\documentclass[twocolumn]{aastex631}

\newcommand\mdot{\dot{M}}
\newcommand\Bv{\langle B_V \rangle}
\newcommand\Rjup{R_\mathrm{Jup}}
\newcommand\Lsun{L_\sun}

\newcommand{\unit}[1]{\ensuremath{\, \mathrm{#1}}}
\newcommand{\resm}{$M=49.5_{-3.5}^{+4.3} \unit{M_\mathrm{Jup}}$}
\newcommand{\resinc}{$i_\star = 114.5^\circ$ $_{-10.0}^{+7.4}$ } 
\newcommand{\updated}[1]{{\color{black}{\bf#1}}}

\usepackage[T1]{fontenc}
\usepackage[utf8]{inputenc} 
\usepackage{comment}
\usepackage{float}
\DeclareUnicodeCharacter{2212}{-}
\begin{document}

\title{Astrometry and Precise Radial Velocities Yield a Complete Orbital Solution for the Nearby Eccentric Brown Dwarf LHS~1610~b}

\correspondingauthor{Evan Fitzmaurice}
\email{exf5296@psu.edu}
\newcommand{\PSUAA}{Department of Astronomy \& Astrophysics, Pennsylvania State University, University Park, PA, 16802, USA}
\newcommand{\PSUCEHW}{Center for Exoplanets and Habitable Worlds, Pennsylvania State University, University Park, PA, 16802, USA}
\newcommand{\PSUICDS}{Institute for Computational and Data Sciences, The Pennsylvania State University, University Park, PA, 16802, USA}
\newcommand{\PSUStats}{Center for Astrostatistics, 525 Davey Laboratory, The Pennsylvania State University, University Park, PA, 16802, USA}
\newcommand{\Princeton}{Department of Astrophysical Sciences, Princeton University, 4 Ivy Lane, Princeton, NJ 08540, USA}
\newcommand{\Goddard}{NASA Goddard Space Flight Center, 8800 Greenbelt Road, Greenbelt, MD 20771, USA}

\newcommand{\ASTRON}{ASTRON, Netherlands Institute for Radio Astronomy, Oude Hoogeveensedijk 4, Dwingeloo, 7991 PD, The Netherlands}
\newcommand{\API}{Anton Pannekoek Institute for Astronomy, University of Amsterdam, 1090 GE Amsterdam, the Netherlands}
\newcommand{\Aarhus}{Stellar Astrophysics Centre, Department of Physics and Astronomy, Aarhus University, Ny Munkegade 120, 8000 Aarhus C, Denmark}
\newcommand{\MCDonald}{McDonald Observatory and Department of Astronomy, The University of Texas at Austin, 2515 Speedway, Austin, TX 78712, USA}
\newcommand{\UTSpace}{Center for Planetary Systems Habitability, The University of Texas at Austin, 2515 Speedway, Austin, TX 78712, USA}
\newcommand{\UCI}{Department of Physics \& Astronomy, The University of California, Irvine, Irvine, CA 92697, USA}
\newcommand{\Queensland}{School of Mathematics \& Physics, University of Queensland, St Lucia, QLD 4072, Australia}
\newcommand{\LeidenObservatory}{Leiden Observatory, Leiden University, PO Box 9513, 2300 RA,
Leiden, The Netherlands}
\newcommand{\PSETI}{Penn State Extraterrestrial Intelligence Center, 525 Davey Laboratory, The Pennsylvania State University, University Park, PA, 16802, USA}
\newcommand{\QueenslandCfA}{Centre for Astrophysics, University of Southern Queensland, West Street, Toowoomba, QLD 4350, Australia}

\author[0000-0003-0199-9699]{Evan Fitzmaurice*}
\affiliation{\PSUAA}
\affiliation{\PSUCEHW}
\affiliation{\PSUICDS}
\affiliation{Institute for Computational and Data Sciences Scholar}

\author[0000-0001-7409-5688]{Guðmundur Stefánsson}
\affil{\API}
\affiliation{NASA Sagan Fellow}
\affiliation{\Princeton}

\author[0000-0002-1486-7188]{Robert D. Kavanagh}
\affil{\ASTRON}
\affil{\API}

\author[0000-0001-9596-7983]{Suvrath Mahadevan}
\affiliation{\PSUAA}
\affiliation{\PSUCEHW}

\author[0000-0003-4835-0619]{Caleb I. Ca\~nas}
\affil{NASA Postdoctoral Fellow}
\affil{\Goddard}

\author[0000-0002-4265-047X]{Joshua N.\ Winn}
\affil{\Princeton}

\author[0000-0003-0149-9678]{Paul Robertson}
\affil{\UCI}

\author[0000-0001-8720-5612]{Joe P.\ Ninan}
\affil{Department of Astronomy and Astrophysics, Tata Institute of Fundamental Research, Homi Bhabha Road, Colaba, Mumbai 400005, India}

\author[0000-0003-1762-8235]{Simon Albrecht}
\affil{\Aarhus}

\author[0000-0002-7167-1819]{J. R. Callingham}
\affil{\LeidenObservatory}
\affil{\ASTRON}

\author[0000-0001-9662-3496]{William D. Cochran}
\affil{\MCDonald}
\affil{\UTSpace}

\author[0000-0003-1439-2781]{Megan Delamer}
\affiliation{\PSUAA}
\affiliation{\PSUCEHW}

\author[0000-0001-6545-639X]{Eric B. Ford}
\affiliation{\PSUAA}
\affiliation{\PSUCEHW}
\affiliation{\PSUICDS}
\affiliation{\PSUStats}

\author[0000-0001-8401-4300]{Shubham Kanodia}
\affil{Earth and Planets Laboratory, Carnegie Institution for Science, 5241 Broad Branch Road, NW, Washington, DC 20015, USA}

\author[0000-0002-9082-6337]{Andrea S.J.\ Lin}
\affil{\PSUAA}
\affil{\PSUCEHW}

\author[0000-0003-2173-0689]{Marcus L. Marcussen}
\affil{\Aarhus}

\author[0000-0003-2595-9114]{Benjamin J. S. Pope}
\affil{\Queensland}
\affil{\QueenslandCfA}

\author[0000-0002-4289-7958]{Lawrence W. Ramsey}
\affil{\PSUAA}
\affil{\PSUCEHW}

\author[0000-0001-8127-5775]{Arpita Roy}
\affil{Astrophysics \& Space Institute, Schmidt Sciences, New York, NY 10011, USA}

\author[0000-0002-0872-181X]{Harish Vedantham}
\affil{\ASTRON}
\affil{Kapteyn Astronomical Institute, University of Groningen, Landleven 12, NL-9747AD Groningen, the Netherlands}

\author[0000-0001-6160-5888]{Jason T.\ Wright}
\affil{\PSUAA}
\affil{\PSUCEHW}
\affil{\PSETI}

\begin{abstract}
The LHS 1610 system consists of a nearby ($d=9.7$ pc) M5 dwarf hosting a candidate brown dwarf companion in a $10.6$ day, eccentric ($e \sim 0.37$) orbit. We confirm this brown dwarf designation and estimate its mass ($ 49.5_{-3.5}^{+4.3}$ $M_{\text{Jup}} $) and inclination ($ 114.5^\circ$ $_{-10.0}^{+7.4}$) by combining discovery radial velocities (RVs) from TRES and new RVs from the Habitable-zone Planet Finder with the available Gaia astrometric two-body solution. We highlight a discrepancy between the measurement of the eccentricity from the Gaia two-body solution ($e=0.52 \pm 0.03$) and the RV only solution ($e=0.3702\pm0.0003$). We discuss possible reasons for this discrepancy, which can be further probed when the Gaia astrometric time series become available as part of Gaia DR4. As a nearby mid M star hosting a massive short-period companion with a well-characterized orbit, LHS 1610 b is a promising target to look for evidence of sub-Alfv\'enic interactions and/or auroral emission at optical and radio wavelengths. LHS 1610 has a flare rate ($0.28\pm 0.07$ flares/day) on the higher-end for its  rotation period ($84 \pm 8$ days), similar to other mid M dwarf systems such as Proxima Cen and YZ Ceti that have recent radio detections compatible with star-planet interactions. While available TESS photometry is insufficient to determine an orbital phase-dependence of the flares, our complete orbital characterization of this system makes it attractive to probe star-companion interactions with additional photometric and radio observations.

\end{abstract}


\section{Introduction}
\label{sec:intro}
The Gaia mission \citep{Gaia} is revolutionizing the field of astrophysics yielding insights into planets, brown dwarfs, and binary stars. The expected detection yield of substellar objects from Gaia---including both exoplanets and brown dwarfs---has been estimated to be thousands to tens of thousands \citep{lattanzi2000,sozzetti2001,perryman2014,holl2022}. Recently, new detections of substellar objects have been enabled through studying proper motion differences between Hipparcos and Gaia, allowing follow-up observations through direct imaging and/or radial velocities to gain insights into brown dwarfs \citep[e.g.,][]{brandt2021,li2023} and giant planets \citep{currie2023}. Previous studies have shown evidence for different formation mechanisms of brown dwarfs and giant planets \citep{chabrier2014}, e.g., through their different eccentricity distributions \citep{bowler2020}. Detailed characterization of the orbital parameter distributions of substellar companions---spanning both brown dwarfs and planetary companions---can yield further insights into how these distinct populations of companions form and evolve.

As part of Gaia Data Release 3 \citep[DR3,][]{GaiaDR3}, 169,277 Gaia two-body solutions were published assuming a single Keplerian model derived from the first 34 months of Gaia observations \citep{Halbwachs2022}. These two-body solutions provide constraints on all orbital elements, including the orbital period, eccentricity, inclination, mass of the companion, and correlation matrices between the parameters assuming a `dark' companion which contributes no light to the photocenter motion measured by Gaia. Most of these solutions are double-star systems, with 1,162 that are likely to be substellar objects analyzed with a dedicated `exoplanet' pipeline \citep{holl2022}. Recent work by \cite{Winn2022} provided an analysis of planet candidates with Gaia two-body solutions, providing a framework to analyze the two-body solutions along with available radial velocity data, which in some cases highlights good agreement with the Gaia solutions, and sometimes inconsistencies. Recent follow-up observations by \citet{marcussen2023} further highlight the importance of ground-based observations to confirm and/or rule out false positive scenarios, such as binaries.

Among different stars, nearby M-dwarfs are particularly suitable for detecting substellar companions with Gaia, as they maximize the likelihood of high-precision orbit and mass determination \citep{sozzetti2014,perryman2014}. Around M-dwarfs, \cite{sozzetti2014} predicted the detection of $\sim$100 giant planets at orbits within $3 \unit{AU}$ within $30 \unit{pc}$, and $\sim$$2,000$ within $100 \unit{pc}$. Such a large sample can place tight constraints on the occurrence rates of substellar companions around M dwarfs, which still remains poorly constrained. Detecting giant planets around M-dwarfs is particularly valuable, as current models do not predict their formation due to the expected inventory of material in the disk being too low \citep[e.g.,][]{miguel2020,burn2021}. Such systems of nearby M stars hosting close-in companions are also prime candidates to search for possible signatures of sub-Alfv\'enic interactions at optical and radio wavelengths to gain insights into the magnetic environments of the orbiting companion \citep[e.g.,][]{Callingham2021,kavanagh23}.

In this paper, we perform a detailed characterization of the LHS 1610 system, the second closest M dwarf with a substellar companion and a Gaia two-body solution. The only closer M dwarf system with a substellar companion and Gaia two-body solution is the planetary system GJ 876 \citep{Rivera2005}. LHS~1610~b was originally detected by \cite{Winters2018}, before Gaia two-body solutions were available. They characterized it as a mid M dwarf system that hosts a likely brown dwarf in a $P=10.6$ day eccentric orbit with a minimum mass of $m\sin i = 44.8 \pm 3.2$ $M_{\text{Jup}}$ obtained with radial velocities from the TRES spectrograph. After the release of the Gaia two-body solutions, the system was highlighted in \cite{arenou2022}, where the Gaia astrometric fit independently confirms the orbit of the brown dwarf, although a joint RV and Gaia two-body solution analysis was not performed. To characterize the LHS 1610 system in further detail, we performed a joint sampling of the Gaia two-body solution along with radial velocities, including the RVs from TRES from \cite{Winters2018} and new precise near-infrared RVs from the Habitable-zone Planet Finder (HPF) spectrograph \citep{mahadevan2012,mahadevan2014} on the 10m Hobby-Eberly Telescope. The combined set of RVs from TRES and HPF constrain a new minimum mass of $44.38\pm0.67$ $M_{\text{Jup}}$. The joint sampling allows us to make a new estimate of the orbital inclination of the companion, and thereby its mass of \resm, confirming that the companion is a brown dwarf. Using available TESS data, we derive a flare rate for LHS 1610 and compare it to other M star flare rates and their rotation periods from \citet{pope2021} and \citet{Medina2020,Medina2022}. We find that LHS 1610 resides at the high end of flare rates for its long rotation period for mid M stars, similar to other mid M stars such as Proxima Centauri, YZ Ceti, and GJ 1151. These stars are promising candidates for sub-Alfv\'enic interactions due to known companions \citep{Faria2022,Anglada-Escude2016,Stock2020,Blanco-Pozo2023} and radio detections \citep{Perez-Torres2021,pineda2023,trigilio2023,vedantham2020gj1151,Callingham2021}. This leads us to speculate if the flaring of LHS 1610 is influenced by interactions with its companion. Since additional investigation is necessary to confirm or rule out that scenario, we assess the feasibility of making such a detection. Due to the large size of the short-period companion around a nearby low mass star, we show that the system is particularly favorable for the detection of possible sub-Alfv\'enic interactions and potential auroral emission from the brown dwarf at radio wavelengths. We provide a framework for future inquiries into star-planet/star-companion interactions using fully characterized orbits via Gaia two-body solutions and ground based radial velocities. 

This paper is structured as follows. In Section \ref{sec:stellarparam}, we discuss the parameters of the host star, and we discuss the observations analyzed in Section \ref{sec:observations}. In Section \ref{sec:methods}, we discuss our modeling of the Gaia two-body solution and the available RVs, and discuss the accompanying results in Section \ref{sec:analysis}. We place the system in context with other brown dwarf systems in Section \ref{sec:comparison}. In Section \ref{sec:interactions}, we discuss our flare analysis of available TESS data, and energetics of possible sub-Alfv\'enic interactions or auroral emission in the system. We conclude with a summary of our findings in Section \ref{sec:summary}.

\section{Stellar Parameters} \label{sec:stellarparam}
To characterize the spectroscopically determined $T_{\mathrm{eff}}$, $\mathrm{[Fe/H]}$, and $\log g$ parameters, we used the \texttt{HPF-SpecMatch} \citep{Stefansson2020} code, which compares an as-observed spectrum with HPF to a library of well-characterized spectra. In doing so, we realized that LHS 1610 is listed in the input library from \cite{Yee2017}, with an effective temperature of $T_{\mathrm{eff}} = 3079 \pm 60 \unit{K}$, metallicity of ${\mathrm{[Fe/H]}}=0.01\pm0.08$, and $\log g = 5.04 \pm 0.06$ as originally determined in \cite{mann2015}. As a test, we removed the LHS 1610 spectrum from the library, and we ran it through the \texttt{HPF-SpecMatch} algorithm, recovering consistent values. We elected to adopt the spectroscopically determined measurements as originally reported in \cite{mann2015}. Additionally, the \texttt{HPF-SpecMatch} analysis further confirms a low projected rotational velocity of $v \sin i < 2 \unit{km/s}$, agreeing with the long rotation period from \cite{Winters2018} of $P=84.3 \unit{days}$ which was securely measured using long-term ground-based photometric monitoring from MEarth \citep{Nutzman2008,Irwin2015}. 

To obtain constraints on the mass, radius, and age of the system, we performed a fit to the Spectral Energy Distribution (SED) of LHS 1610 using available literature magnitudes of the star using the \texttt{EXOFASTv2} \citep{Eastman2019} code and MESA Isochrones and Stellar Tracks (MIST;  \citealt{choi2016,dotter2016}) isochrones. As an input for the SED fit, we used informative priors on the spectroscopically determined $T_{\mathrm{eff}}$, $\mathrm{[Fe/H]}$, and $\log g$ parameters as listed above. In doing so, we obtain a mass of $M = 0.167_{-0.015}^{+0.014} \unit{M_\odot}$, and a radius of $R=0.2007_{-0.0063}^{+0.0071} \unit{R_\odot}$. As a separate constraint on the stellar mass, we used the M-K relation from \cite{mann2019}, where we find a stellar mass of $M = 0.1671 \pm 0.0041 M_\odot$. This agrees with the stellar mass from the SED fit, but is more precise. We elected to adopt the mass from the M-K relation, as the relation has been tightly calibrated for mid-to-late M-dwarfs. Table \ref{tab:stellarparam} summarizes our adopted stellar parameters.

\begin{deluxetable}{llcc}
\tablecaption{Summary of stellar parameters used in this work. \label{tab:stellarparam}}
\tabletypesize{\scriptsize}
\tablehead{\colhead{Parameter} &  \colhead{Description}     & \colhead{Value}                               & \colhead{Reference}}
\startdata      
SpT                            &  Spectral Type             & M5                                            & (1)         \\
$T_{\mathrm{eff}}$             &  Effective Temperature     & $3079 \pm 60 \unit{K}$                        & (2)         \\ 
$\mathrm{[Fe/H]}$              &  Metallicity               & $0.01 \pm 0.08$                               & (2)         \\ 
$\log(g)$                      &  Surface gravity (cgs)     & $5.04 \pm 0.06$                               & (2)         \\ 
$R_*$                          &  Radius                    & $0.2007_{-0.0063}^{+0.0071} \unit{R_{\odot}}$ & This Work   \\ 
Age                            &  Age                       & $7.0_{-4.7}^{+4.5} \unit{Gyr}$                & This Work   \\ 
$M_*$                          &  Mass                      & $0.1671 \pm 0.0041 \unit{M_{\odot}}$          & This Work   \\ 
$RV$                           &  Systemic RV               & $43.1 \pm 0.1 \unit{km\ s^{-1}}$              & This work   \\ 
$d$                            &  Distance                  & $9.6625_{-0.0088}^{+0.0090} \unit{pc}$        & (3)         \\ 
$\varpi$                       &  Parallax                  & $103.879_{-0.023}^{+0.023} \unit{mas}$        & (4)         \\
$P_{\mathrm{rot}}$             &  Rotation Period           & $84.3 \pm 8 \unit{days}$                      & (1)         \\ 
$v \sin i$                     &  Rotational Velocity       & $<2 \unit{km/s}$                              & This Work   \\ 
$R_{\mathrm{flare}}$           &  Flare Rate$^{\text{a}}$           & $0.28 \pm 0.07 \unit{day^{-1}}$               & This Work   \\
$\ln{R_{31.5}}$                &  "High Energy" Flare Rate$^{\text{b}}$         & $-2.51 \pm 0.45 \unit{day^{-1}}$              & This Work   \\
\enddata
\tablenotetext{}{References are: (1) \cite{Winters2018}, (2) \cite{mann2015} (3) \cite{bailer-jones2018}, (4) Gaia.}
\tablenotetext{}{\cite{Winters2018} report a rotation period of 84.3 days with a 5-10\% error. We adopt a 10\% rotation period error.}
\tablenotetext{}{ $^{\text{a}}$Derived following the methodology in \cite{pope2021}.}
\tablenotetext{}{ $^{\text{b}}$Derived following the methodology in \cite{Medina2020,Medina2022}.}
\end{deluxetable}

\section{Observations}
\label{sec:observations}

\subsection{TRES Optical Radial Velocities}
We use RVs of LHS 1610 from \cite{Winters2018} which used the Tillinghast Reflector Echelle Spectrograph (TRES). There are a total of 13 RVs that have a median RV uncertainty of 28.3 m/s and span 39 days. The spectra were taken with 900 second exposures in good conditions, and longer in poor conditions. The medium fiber was used with a resolving power of R $\sim$ 44,000. The signal-to-noise ratio (SNR) was 15 per pixel at 7150 Angstroms. The RVs were extracted using the pipeline described in \cite{Buchhave2010}, and are shown in Panel A of Figure \ref{fig:jointfit}.

\subsection{HPF Near-infrared Radial Velocities}
We acquired precise RVs from the spectra of LHS 1610 using the Habitable-zone Planet Finder spectrograph \citep[HPF]{mahadevan2012,mahadevan2014}. HPF is a fiber-fed near-infrared (NIR) spectrograph on the 10 m Hobby-Eberly Telescope \citep[HET,][]{Ramsey1998,Hill2021,Shetrone2007} at McDonald Observatory in Texas, covering the $z$, $Y$, and $J$ bands from 810 to 1280 nm at a resolving power of $R \sim 55,000$. To enable precise RVs in the NIR, the HPF is temperature-stabilized at the milli-Kelvin level \citep{Stefansson2016}. We extracted the HPF RVs using a modified version of the SpEctrum Radial Velocity AnaLyzer \citep[SERVAL;][]{zechmeister2018}, adapted for HPF following \cite{metcalf2019} and \cite{Stefansson2020,stefansson2023}.

In total we obtained 6 HPF observations, which have a median SNR of 158 at 1 micron and a median RV uncertainty of 4.7 m/s. The RVs span 528 days, significantly expanding the total observational baseline. Three of the RV points are the average of two 969 second exposures taken in the same night. Two of the RV points are singular 969 second exposures and the final RV point comes from a spectrum taken with an exposure time of 191 seconds. The RVs are shown in Panel B of Figure \ref{fig:jointfit}, and listed in Table \ref{tab:appendix_rvs}.

\subsection{Gaia Two-Body Solution}
As part of Gaia DR3, fits indicating two bodies are reported in the Gaia Archive under the \texttt{nss\_two\_body\_orbits}). These two-body fits contain best-fit results for parameters along with a correlation matrix quantifying the correlation between the parameters. We convert the correlation matrix to a covariance matrix using \texttt{nsstools} \footnote{\url{https://www.cosmos.esa.int/web/gaia/dr3-nss-tools}} \citep{Halbwachs2022}. There were 445 astrometric CCD observations used for the Gaia two-body solution fit of LHS 1610 as provided by the Gaia team. Relevant Campbell elements from the Gaia two-body solution are listed in column 4 of Table \ref{tab:results}. 

\subsection{TESS Photometry}\label{subsec:TESS}
LHS 1610 was observed by the Transiting Exoplanet Survey Satellite (TESS) \citep{Ricker2015} in Sectors 42 (2021 August 20-2021 September 16), 43 (2021 September 16-2021 October 12), and 44 (2021 October 12-2021 November 6). In the TESS Input Catalog \citep{Stassun2018,Stassun2019}, LHS 1610 is listed as TIC 242941982. LHS 1610 shows flares in each TESS Sector; the TESS Sectors with highlighted flares are shown in Appendix \ref{sec:appendix_flares}. We determine the flare rate and put this rate in context with other M dwarfs in Section \ref{sec:flaring}

\section{Modeling of Astrometry and Radial Velocities}\label{sec:methods}
To characterize the system, we compare the results from three different methods: 1) the Gaia two-body solution, 2) a fit of the radial velocities (`RV-only fit'), and 3) a joint sampling of both the Gaia two-body solution and the available RVs. For the joint sampling, we broadly follow the methodology outlined in \cite{Winn2022}. The three models and values are further described below.

\subsection{Gaia Two-Body Solution}\label{subsec:astrometryonly}
The two-body solution from the Gaia DR3 \texttt{nss\_two\_body\_orbits} table yields constraints on the following parameters:
\begin{equation}
 A, B, F, G, e, P, t_p, \varpi,
\end{equation}
where $e$ is eccentricity, $P$ is period, $\varpi$ is the parallax, and $t_p$ is the periastron time referenced to epoch 2016.0 (JD 2,457,389.0). $A$, $B$, $F$, and $G$ are the Thiele-Innes coefficients:
\begin{equation}
A = a_0(\text{cos}\omega \: \text{cos}\Omega - \text{sin}\omega \:\text{sin}\Omega \:\text{cos}i),
\end{equation}
\begin{equation}
B = a_0(\text{cos}\omega \:\text{sin}\Omega + \text{sin}\omega \:\text{cos}\Omega \:\text{cos}i),
\end{equation}
\begin{equation}
F = -a_0(\text{sin}\omega \:\text{cos}\Omega + \text{cos}\omega \:\text{sin}\Omega \:\text{cos}i),
\end{equation}
\begin{equation}
G = -a_0(\text{sin}\omega \:\text{sin}\Omega - \text{cos}\omega \:\text{cos}\Omega \:\text{cos}i)
\end{equation}
where $a_0$ is the semimajor axis of the photocenter converted to milliarcseconds by multiplying by the parallax, $\omega$ is the argument of periastron, $\Omega$ is the longitude of the ascending node, and $i$ is the inclination. We use the covariance matrix and \texttt{nsstools} to use the Thiele-Innes coefficients to yield constraints on $\omega$, $\Omega$, $i$, and $a_0$. \cite{Halbwachs2022} discuss the ranges of these elements and their physical interpretation from the Gaia two-body solution fits. The astrometric fit uniquely constrains the orbital inclination to the physical motion of the orbit, where orbital inclinations between [0,$\frac{\pi}{2}$] indicate a counterclockwise orbit, while values between [$\frac{\pi}{2}$,$\pi$] indicate a clockwise orbit. Due to a degeneracy of $\pi$ in $\Omega$ and $\omega$, the Gaia astrometric orbit fits will have two equivalent possible solutions, and as noted by \cite{Halbwachs2022} the solution provided in the Gaia two-body solution table is the solution where $\Omega$ is between [0,$\pi$] and $\omega$ is between [0,2$\pi$]. 

\subsection{RV-only fit}\label{subsec:doppleronly}
For the RV-only fit, we use the following as parameters in the fit:
\begin{equation}\label{eq:rv-params}
\text{\emph{P}}, t_p, e, \omega, K, \gamma,
\end{equation}
where $K$ is the radial velocity semi-amplitude, and $\gamma$ is the radial velocity offset for the spectrograph combined with the stellar RV offset. We fit a new $\gamma$ for each individual spectrograph. We compute the Keplerian RV model using the \texttt{radvel} code \citep{fulton2018}. RVs alone allow us to uniquely constrain the value of $\omega$ but not the inclination, meaning we cannot determine the true mass of the secondary, only its minimum mass.

The Doppler likelihood function is:
\begin{equation}\label{eq:RVonlylikelihood}
\mathcal{L}_v = \prod_{i = 1}^{N} \frac{1}{\sqrt{2\pi(\sigma^2_{v,i})}} \text{exp} \left[-\frac{(v_i - v_{i,\text{calc}})^2}{2(\sigma^2_{v,i})} \right]
\end{equation}
where $v_i$ is the $i$-th RV data point, $\sigma_{v,i}$ is the associated uncertainty, and $v_{i,\text{calc}}$ is the $i$-th model calculated RV. In practice, we take the log of Equation \ref{eq:RVonlylikelihood} so that we can sum the log value of every $i$-th step. 

To fit the RVs, we use the differential evolution package \texttt{PyDE} \citep{pyde} to determine a global maximum-likelihood solution of a Keplerian RV model to the RV observations. We then initialize 100 Markov-Chain Monte Carlo (MCMC) walkers around the global maximum-likelihood solution to perform MCMC sampling of the parameter posteriors using the \texttt{emcee} package \citep{dfm2013}. We ran the walkers for 35,000 steps. After removing a burn-in of 2,000 chains, we assess the convergence of the chains with two metrics. First, we compute the Gelman-Rubin statistic and confirm its value for each parameter is within 1\% of unity. This statistic can be unreliable if the chains are not independent \citep[see e.g., discussion in][]{Hogg2018}. Therefore, we additionally computed the maximum autocorrelation timescale and adjust the number of chains in the MCMC to be longer than 50 times this value to ensure a sufficient number of independent samples\footnote{\url{https://emcee.readthedocs.io/en/latest/tutorials/autocorr/}}. We find the mean autocorrelation time to be 78 and the maximum to be 109, so from these steps, along with visual inspections of the chains, we conclude that the chains are well-mixed.

\subsection{Joint Astrometry and RV sampling}\label{subsec:jointfit}
To jointly sample the Gaia astrometric covariance matrix and the available RVs, we use the following parameters,
\begin{equation}
M_*,m_2,e,\omega,\cos i,\Omega,\text{\emph{P}}, t_p, \varpi, \gamma, \varepsilon, \sigma_{\text{scale}},
\end{equation}
where $M_*$ is the stellar mass, $m_2$ is the secondary mass, $\cos i$ is the cosine of the inclination, $\varepsilon$ is the flux ratio between the companion and star, and $\sigma_{\text{scale}}$ is a scaling factor that acts as a jitter term for the two-body solution and is discussed below. All of the Gaia parameters are relative to the photocenter since Gaia measures the center of light between the star and any potentially unresolved companions. $\varepsilon$ is involved when computing the $a_0$ for a given set of parameters:

\begin{equation}\label{eq:Winneq17}
    \frac{a_0}{\varpi} = \left[G(M_* + m_2)\right]^{\frac{1}{3}} \left(\frac{P}{2\pi}\right)^{\frac{2}{3}} \left(\frac{m_2}{M_* + m_2} - \frac{\varepsilon}{1+\varepsilon}\right).
\end{equation}

Like the RV only fit, we fit individual RV offsets for HPF and TRES.  

To sample the Gaia covariance matrix, we modify the Gaia likelihood function from \cite{Winn2022}:
\begin{equation}
\mathcal{L}_g = \frac{1}{\sqrt{(2\pi)^8|\text{det}\mathcal{C}_{\text{scale}}|}} \text{exp} \left[-\frac{1}{2} (\Theta^{\text{T}} \mathcal{C}_{\text{scale}}^{-1} \Theta ) \right],
\end{equation}
where $\mathcal{C}_{\text{scale}}$ is the covariance matrix including a multiplicative factor, $\sigma_{\text{scale}}$ used to optionally scale the full covariance matrix, where

\begin{equation}
\mathcal{C}_{\text{scale}} = \sigma_{\text{scale}}^2 \times \mathcal{C},
\end{equation}
where $\mathcal{C}$ is the original covariance matrix from the Gaia two-body solution.
$\sigma_{\text{scale}}$ is squared in the above equation to interpret the scaling factor as a multiplicative uncertainty scaling factor rather than a multiplicative variance factor. As $\sigma_{\text{scale}}$ uniformly scales the full covariance matrix, this has the effect of increasing the uncertainties between the different parameters, while keeping the relative correlations between the parameters the same.

$\Theta$ is the "Gaia deviation vector" which is an 8-column vector of the differences between the Gaia reported value and the calculated value for the following parameters:
\begin{equation}
 A, B, F, G, e, P, t_p, \varpi.
\end{equation}
When computing the RV model to determine the RV likelihood in the joint sampling, we calculate the semi-amplitude according to Equation 15 from \citet{Winn2022}:
\begin{equation}\label{eq:K_from_a0}
K = \frac{2\pi}{P} \frac{(a_0/\varpi) \sqrt{1-\text{cos}^2 i}}{\sqrt{1-e^2}}
\end{equation}
while $a_0$ is calculated using Equation \ref{eq:Winneq17}.

The total likelihood for the joint sampling is therefore given by:
\begin{equation}
\label{eq:joint_total_likelihood}
\log(\mathcal{L}_{\text{Total}}) = \log(\mathcal{L}_{g} ) + \log(\mathcal{L}_{v} ).
\end{equation}

Similar to the RV-only analysis, we use \texttt{PyDE} to find a global maximum-likelihood solution, after which we use \texttt{emcee} to perform MCMC sampling of the posteriors. For the analysis, we ran two runs, a joint sampling where $\sigma_{\mathrm{scale}}$ is fixed to 1, i.e., using the covariance matrix as is, and another run where $\sigma_{\mathrm{scale}}$ is allowed to float, where the latter was done to help account for unexplained discrepancies seen between the RV and the Gaia two body solution. 

For the former sampling where $\sigma_{\mathrm{scale}}=1$, we initialize 100 walkers and run those walkers for 45,000 steps. We removed the first 2,500 chains as burn-in chains. Our Gelman-Rubin statistics are all within 1\% of unity. The mean autocorrelation timescale is 178, while the maximum is 282, meaning our chains are well mixed.

For the second sampling run where $\sigma_{\mathrm{scale}}$ is allowed to float, we initialize 100 walkers and run the walkers for 3,250,000 steps. We removed the first 100,000 steps as burn-in. The Gelman-Rubin statistic are again within 1\% of unity for all parameters. The maximum autocorrelation length is 56,730 for the parameters which is less than 50 times the length of the chains minus the burn-in, suggesting that our chains are well mixed. As mentioned in Sections \ref{subsec:doppleronly} and \ref{subsec:astrometryonly}, the RVs uniquely constrain $\omega$ but not the inclination nor $\Omega$, while the astrometry uniquely constrains inclination, but has the $\pi$ degeneracy for $\omega$ and $\Omega$. By jointly sampling the RVs and the astrometric solution, we can constrain all three values and break the degeneracies.

\section{Results}\label{sec:analysis}
\begin{figure*}[t!]
\begin{center}
\includegraphics[width=\textwidth]{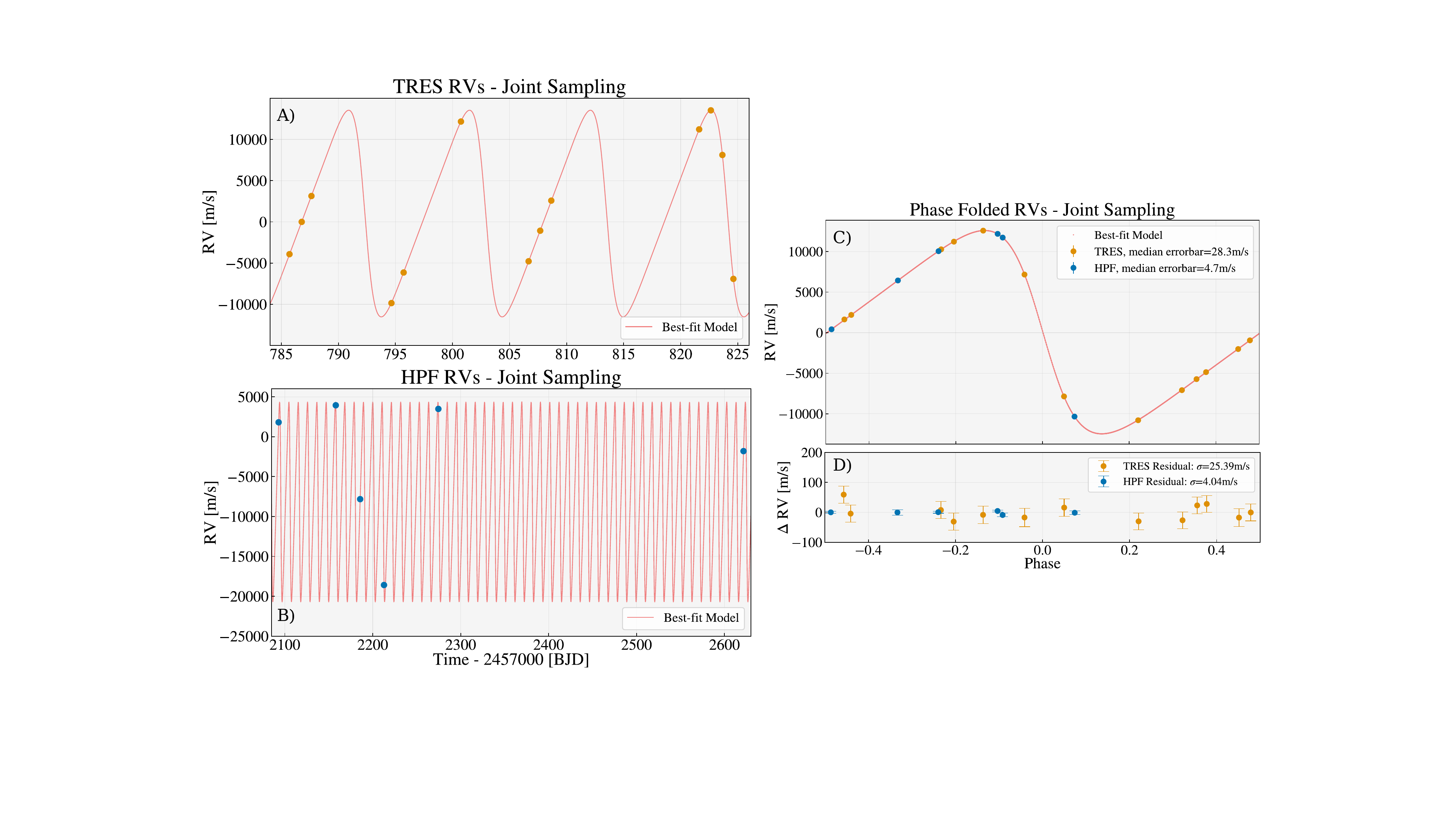}
\vspace{-0.8cm}
\end{center}
\caption{Visualization of joint Gaia-RV sampling with inflation.  \textbf{A:} RVs using TRES from \cite{Winters2018}, with our best-fit model laid over the points. \textbf{B:} Our new RVs using HPF, with the best fit model in red. This shows the significant increase in the observing baseline provided by the additional HPF RVs which span more than 500 days. \textbf{C and D:} Phase folded RVs and residuals using the parameters produced by the joint sampling. Both sets of RVs are in agreement with the best-fit model. Errors in panels A, B, and C are smaller than the marker size. The RVs are available in a machine readable table with the manuscript.}
\label{fig:jointfit}
\end{figure*}

Below we discuss the results from the three different methodologies we used to constrain the orbital parameters of LHS 1610 b: from the available RVs (RV-only), the Gaia two body solution (Gaia only), and a joint sampling of both the RVs and the Gaia two-body solution with and without a $\sigma_{\mathrm{scale}}$ factor for the Gaia covariance matrix. Table \ref{tab:results} summarizes the resulting posteriors from the different methodologies. The results from our final adopted joint-sampling model are graphically summarized in Figure \ref{fig:jointfit}.

\subsection{RV-only}
First, we compare our results in Table \ref{tab:results} from the RV-only fit to the values in \cite{Winters2018}. The values we obtain are consistent with the values reported in \cite{Winters2018} which used only RVs from the TRES spectrograph. The additional HPF RVs allow us to significantly improve the precision on multiple parameters including decreasing the orbital period uncertainty by a factor of 140, and the eccentricity is more tightly constrained as $e = 0.37019 \pm 0.0003$ compared to $e = 0.36942 \pm  0.00093$ from \cite{Winters2018}. We experimented running fits with individual RV jitter parameters for TRES and HPF that had Jeffreys priors from 0.1 to 100 m/s. This returned Keplerian parameter constraints consistent with the fits with no jitter. The jitter estimates were $1.6^{+9.3}_{-1.3}$ m/s and $4.4^{+8.7}_{-3.9}$ m/s for TRES and HPF, respectively. The modes of the distributions of the jitter values are 0 m/s. From this, we interpret that the RV uncertainties provide an accurate estimate of the total uncertainties and we elect to list the posterior results from the RV-only fit without the jitter values. The RVs from TRES and HPF yield a minimum mass constraint of $m_2 \sin(i) = 44.38\pm0.67$ $M_{\text{Jup}}$.

\subsection{RV-only vs.~Gaia Two-Body Solution}
Second, we compare our RV-only results to the results from the Gaia two-body solution (third column in Table \ref{tab:results}). We see that the time of periastrion is consistent and the orbital period constraint is broadly in agreement, although with the radial velocity determined period being much more precise ($P = 10.5885\pm0.0013$ days from the Gaia two body solution, compared to $P = 10.594724\pm0.000016$ from the RVs). Additionally, we see that the value for $\omega$ between the RV-only and the Gaia Two-body solution differ by $\sim180^\circ$. This is a result of the $\pi$ degeneracy in $\omega$ in the astrometric two-body solution fits (see Section \ref{subsec:astrometryonly} and Appendix B of \citealt{Halbwachs2022}). An important note is that the inclination for Gaia sets the direction of the orbit. Since the inclination is between [$\frac{\pi}{2}$,$\pi$] the system orbits in a clockwise direction as observed from Earth.

Importantly, Table \ref{tab:results} highlights a discrepancy between the eccentricity of the two solutions, where the Gaia solution suggests an eccentricity of $e=0.524\pm0.027$, while the RV fit suggests an eccentricity of $e=0.37019\pm0.00003$. From the quality of the RVs, the RV derived eccentricity is reliable, suggesting possible issues with the Gaia two body solution.

A few possibilities could explain this discrepancy. First, a third body may be present in the system that could be biasing the astrometry. However, we deem this unlikely as the RV residuals do not exhibit additional structure or trends from the single companion Keplerian fits.

Another possibility could be that the secondary companion is contributing secondary light, breaking the assumption of a dark companion in the Gaia two body solution. To check if this assumption is warranted in the LHS 1610 system, we estimated the flux ratio, $\varepsilon$, between the brown dwarf and host star in the Gaia bandpass. Using the Sonora-Bobcat spectral templates \citep{marley2021} for the brown dwarf, PHOENIX stellar spectral templates \citep{Husser2013} for the star, and accounting for the transmission curve across the Gaia bandpass, we estimate the flux ratio in the Gaia bandpass (320-1100 nm) to be negligible ($<10^{-6}$). As such, we do not expect light emitted from the brown dwarf to impact the Gaia solution.

The Gaia two-body solution and the associated covariance matrix may not be accurately depicting the shapes of the posteriors of the orbital parameters, but highly structured and/or other non-linear covariances between different parameters would likely not be accurately estimated using the sampling methodology we used (see further discussion in \citealt{Winn2022} and \citealt{marcussen2023}). 

Additionally, we note that astrometric photocenter motion dominated by motion along one axis in the sky plane can result in biases of the orbital fit, including a bias towards close-to edge-on inclinations which in turn would impact the derived eccentricity. Based on the orbital parameters, the orientation of LHS 1610 b's orbit on the sky suggests motion primarily along the declination axis. This could help explain the eccentricity discrepancy given the nearly edge-on inclination of the Gaia two-body solution. 

Lastly, the nominal scanning law that dictates how Gaia observes the sky and the resulting projection of the orbit on the detector may also be at play. For more detail, see \citet{Holl2023} which discusses how the along-scan and across-scan angles of Gaia can impact the derived orbital parameters.

\subsection{Joint RV+Gaia sampling}
\label{subsec:no-inflation}
Given the availability of a Gaia two-body solution for LHS 1610b, we performed a joint sampling of both the Gaia covariance matrix and the radial velocities. Due to the discrepancies between the RV-only solution and the Gaia two-body solution, we consider two separate joint sampling models. First a joint sampling where $\sigma_{\mathrm{scale}}=1$, and second, a joint sampling where we let $\sigma_{\mathrm{scale}}$ be a fit parameter.

\subsubsection{Joint sampling with $\sigma_{\mathrm{scale}}=1$}
\label{subsec:jointsamplingA}
First, we performed a joint sampling of the two-component likelihood function in Equation \ref{eq:joint_total_likelihood}, which includes a likelihood of the Gaia covariance matrix with $\sigma_{\mathrm{scale}}=1$ along with the RV likelihood. The results from this joint sampling are summarized in column 4 of Table \ref{tab:results}. From this joint sampling, we find that the orbital period and the eccentricity converges on the RV-derived values with the eccentricity estimated as $e=0.37004\pm0.000030$. 

As noted previously, the Gaia astrometric fits have two degenerate solutions for $\omega$ and $\Omega$, and the Gaia two-body solution is the solution where $\Omega$ is bound to [0,$\pi$] and $\omega$ to [0,2$\pi$]. The RVs break the $\pi$ degeneracy and informs us of the correct solution: where $\Omega = -14.9 \pm 0.82^\circ$ and $\omega = 89.31 \pm 0.14^\circ$.

This joint sampling results in an inclination of $i=117.19^{\circ}$ $_{-0.91}^{+0.88}$ and a corresponding secondary mass of $50.94\pm0.9 \unit{M_{\mathrm{Jup}}}$. This inclination is $24^\circ$ greater than the value expected by the Gaia two-body solution of $i=92.8 \pm 1.9^\circ$. This discrepancy highlights that the uncertainties from this joint sampling are underestimated. 

To further assess the quality of the joint sampling, we follow \cite{Winn2022} and we consider the Z-scores of the jointly constrained parameters $A$, $B$, $F$, $G$, $e$, $P$, and $t_{\mathrm{p}}$, where the Z-score for a given parameter is calculated as
\begin{equation}
    \text{Z-score} = \frac{x_1 - x_2}{\sigma_{x_2}}
\end{equation}
where $x_1$ and $x_2$ are two different measurements of the same parameter and $\sigma_{x_2}$ is the uncertainty on the $x_2$ measurement. Here we define the Z-scores such that $x_1$ is the median joint-sampling posterior value, $x_2$ is the median Gaia-only posterior value, and  $\sigma_{x_2}$ is the average of the upper and lower uncertainties on the Gaia-only posterior value. In the case where $\sigma_{\mathrm{scale}}$ is a fit parameter, the original Gaia-only uncertainties are multiplied by the median fit value of $\sigma_{\mathrm{scale}}$ before computing $\sigma_{x_2}$.

\begin{figure}[t!]
\begin{center}
\includegraphics[width=0.99\columnwidth]{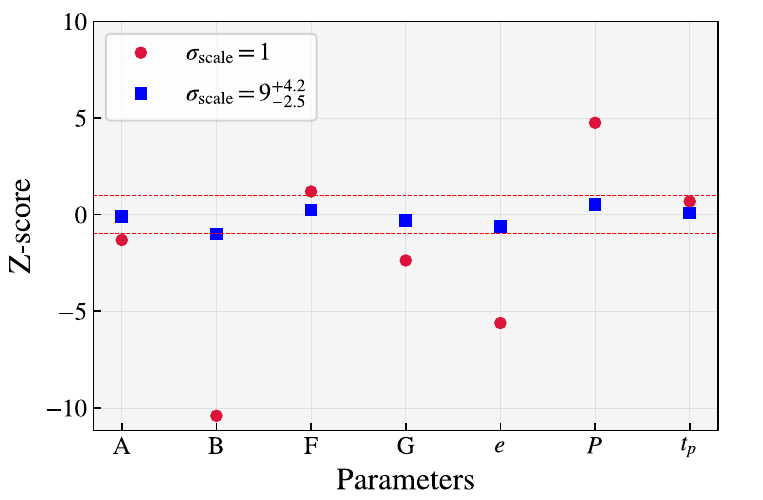}
\vspace{-0.8cm}
\end{center}
\caption{Comparison of Z-score values for the two joint-sampling methodologies we considered. The joint sampling Z-scores where we fix $\sigma_{\mathrm{scale}}=1$ are shown with red circles. Same as before but where we let $\sigma_{\mathrm{scale}}$ be a fit parameter in the MCMC sampling, shown with blue squares. The red dashed lines show Z-score values of $\pm1$. The former sampling highlights inconsistencies (absolute Z-score values larger than 1) for the parameters $e$, $P$ and $B$. From the latter sampling, we see that through letting $\sigma_{\mathrm{scale}}$ inflate the Gaia covariance matrix, this correspondingly increases the uncertainties in the derived parameters (see Table \ref{tab:results}) and thus results in Z-scores that are in better agreement (absolute Z-scores $<1$).}
\label{fig:zscore}
\end{figure}

Figure \ref{fig:zscore} shows the Z-scores for the Thiele-Innes coefficients $A$, $B$, $F$, $G$, along with $e$, $P$, and $t_{\mathrm{p}}$. From Figure \ref{fig:zscore}, we see that the Z-scores highlight inconsistencies in particular between the eccentricity $e$, orbital period $P$ and the $B$ coefficient. From these inconsistencies, we caution against interpreting these values as the final orbital parameters for LHS 1610b.

\begin{deluxetable*}{lcccc}
\centering
\tablecaption{Results from the methods considered in this work. In the Gaia+RV jointly sampled constraints we fix the flux ratio parameter at $\varepsilon=0$. The $\sigma_{\mathrm{scale}}=1$ sampling shows evidence of inconsistencies and likely has underestimated uncertainties. To better account for these inconsistencies, we adopt the jointly sampled Gaia+RV values where we let the $\sigma_{\mathrm{scale}}$ parameter float which results in more conservative uncertainties. We note that these values need to be revisited when the Gaia DR4 astrometric time series become available.}
\label{tab:results}.
\tabletypesize{\scriptsize}
\tablehead{\colhead{Parameter}  &  \colhead{RV-only Fit}          & \colhead{Gaia Two-Body Solution}  & \colhead{Gaia+RV} & Gaia+RV\\
\colhead{}  &  \colhead{}          & \colhead{}  & \colhead{$\sigma_{\mathrm{scale}}\equiv1$$^\dagger$} & $\sigma_{\mathrm{scale}}>1$ (adopted)}
\startdata
M$_*$ ($M_{\odot}$)     & -                                 & -                                   & $0.1672^{+0.0041}_{-0.0041}$        & $0.1672^{+0.0041}_{-0.0041}$      \\
m$_2$ ($M_{\rm Jup}$)   & -                                 & -                                   & $50.94^{+0.89}_{-0.90}$             & $49.5^{+4.3}_{-3.5}$ \\
$\cos i$                & -                                 & -                                   & $-0.457^{+0.014}_{-0.014}$          & $-0.41^{+0.17}_{-0.17}$ \\
i ($^\circ$)            & -                                 & $92.8^{+1.9}_{-1.9}$                & $117.19^{+0.88}_{-0.91}$            & $114.5^{+7.4}_{-10.0}$  \\
K (m/s)                 & $12534.6^{+9.7}_{-9.6}$           & -                                   & $12540.7^{+9.7}_{-9.7}$             & $12534.7^{+9.8}_{-9.6}$ \\
e                       & $0.37019^{+0.00030}_{-0.00030}$     & $0.524^{+0.027}_{-0.027}$         & $0.37004^{+0.00030}_{-0.00030}$    & $0.37018^{+0.00030}_{-0.00030}$ \\
$\omega$ (degs)         & $89.22^{+0.14}_{-0.14}$           & $271.4^{+2.9}_{-2.9}$               & $89.31^{+0.14}_{-0.14}$             & $89.22^{+0.14}_{-0.14}$ \\
$\Omega$ (degs)         & -                                 & $162.6^{+1.5}_{-1.5}$               & $-14.9^{+0.82}_{-0.81}$             & $-14.6^{+8.3}_{-7.7}$\\
t$_{\mathrm{peri}}$ (days)$^{\star}$ & $0.7107^{+0.0031}_{-0.0031}$      & $0.61^{+0.15}_{-0.15}$              & $0.7119^{+0.0031}_{-0.0031}$        & $0.7107^{+0.0032}_{-0.0031}$ \\
P (days)                & $10.594724^{+0.000016}_{-0.000016}$ & $10.5885^{+0.0013}_{-0.0013}$     & $10.594733^{+0.000016}_{-0.000016}$ & $10.59472^{+0.000020}_{-0.000016}$ \\
$\varpi$ (mas)                & -                                 & -                              & $103.881^{+0.023}_{-0.023}$          & $103.879^{+0.023}_{-0.023}$ \\
$\gamma_{\mathrm{TRES}}$ (m/s)   & $945.2^{+8.2}_{-8.2}$             & -                                   & $944.0^{+8.2}_{-8.2}$             & $945.2^{+8.2}_{-8.2}$ \\
$\gamma_{\mathrm{HPF}}$ (m/s)    & $−8244.1^{+3.2}_{-3.2}$           & -                                   & $-8245.7^{+3.2}_{-3.2}$           & $-8244.1^{+3.2}_{-3.2}$  \\
$\sigma_{\text{scale}}$ & -                                 & -                                   & $\equiv1$                                  & $9.0^{+4.2}_{-2.5}$ \\
$a_0$ (mas)                   & -                                 & $1.391^{+0.037}_{-0.037}$     & $1.325^{+0.011}_{-0.011}$         & $1.294^{+0.093}_{-0.078}$ \\
A (mas)                 & -                                 & $-0.053^{+0.067}_{-0.067}$         & $-0.140^{+0.009}_{-0.010}$          & $-0.108^{+0.086}_{-0.089}$ \\
B (mas)                 & -                                 & $-0.055^{+0.051}_{-0.051}$          & $-0.589^{+0.023}_{-0.023}$       & $-0.52^{+0.23}_{-0.19}$ \\
F (mas)                 & -                                 & $-1.327^{+0.037}_{-0.037}$          & $-1.282^{+0.012}_{-0.012}$       & $-1.246^{+0.085}_{-0.098}$ \\
G (mas)                 & -                                 & $0.418^{+0.035}_{-0.035}$           & $0.334^{+0.018}_{-0.018}$        & $0.32^{+0.16}_{-0.19}$ \\
\enddata
\tablenotetext{}{$^\dagger$ Uncertainties likely underestimated. See Section \ref{subsec:jointsamplingA} for further discussion.}
\tablenotetext{}{$^\star$For the periastron time, we follow the Gaia convention where the periastron time is $t_{\mathrm{peri}} = 2457389.0 + t_p$, where $t_p$ is the value listed in the table above.}
\end{deluxetable*}


\subsubsection{Joint sampling with $\sigma_{\mathrm{scale}}$ as a free parameter}
\label{sec:adopted parameters}
To account for the discrepancies between the Gaia two body solution and the radial velocity fit, we performed a second sampling of the joint likelihood function where we let $\sigma_{\mathrm{scale}}$ be a free parameter in the MCMC sampling. This scales the uncertainties in all of the parameters in the Gaia covariance matrix, while keeping the relative covariances.

The results of this are shown in the fifth column of Table \ref{tab:results} and are graphically summarized in Figure \ref{fig:jointfit}. In panels C and D, we see that both the optical TRES and the NIR HPF RVs fully agree on the RV orbit showing RV residuals with no visually apparent residual structure, suggesting a good fit. We also see that the resulting parameters that are directly constrained by the RVs all fully agree with the RV-only Fit in Table \ref{tab:results}. 

From Table \ref{tab:results}, we see that $\sigma_{\mathrm{scale}}=9.0^{+4.2}_{-2.5}$, suggesting that a significant scaling of the Gaia covariance matrix is needed to self-consistently jointly sample the Gaia covarience matrix and the available RVs. As expected, this value of $\sigma_{\mathrm{scale}}$ results in larger uncertainties in the astrometric parameters, including the inclination of \resinc and correspondingly the mass estimate of LHS 1610 b of \resm. Despite the large scaling factor of $\sigma_{\mathrm{scale}}=9.0^{+4.2}_{-2.5}$, the fractional uncertainty on the mass of LHS~1610~b is still at the $\sim7-9\%$ level, sufficient to conclude that the object is a brown dwarf.

Similar to Section \ref{subsec:jointsamplingA}, we assess the quality of the astrometric parameters through investigating the resulting Z-scores which are highlighted in Figure \ref{fig:zscore}. Figure \ref{fig:zscore} shows that increasing the uncertainties with $\sigma_{\mathrm{scale}}$ results in Z-scores which all have absolute values $<1$ suggesting that the sampled values are now in much better agreement given the updated uncertainties. Given the updated Z-scores and the agreement in the RV fit, we adopt these values as the formal parameters for LHS~1610~b.

We note that the adopted parameters assume that the joint sampling methodology is able to give a good description of the parameters. As mentioned previously, we note that without having access to the underlying astrometric Gaia time-series, it is impossible to discern the exact cause of the underlying discrepancies. We urge the community to revisit this system and potentially other similar systems like it, to test the validity of this approach and the consistency of the final derived parameters.


\section{Comparison to Other Brown Dwarf M dwarf Systems} \label{sec:comparison}

\begin{figure*}[t!]
\begin{center}
\includegraphics[width=0.93\textwidth]{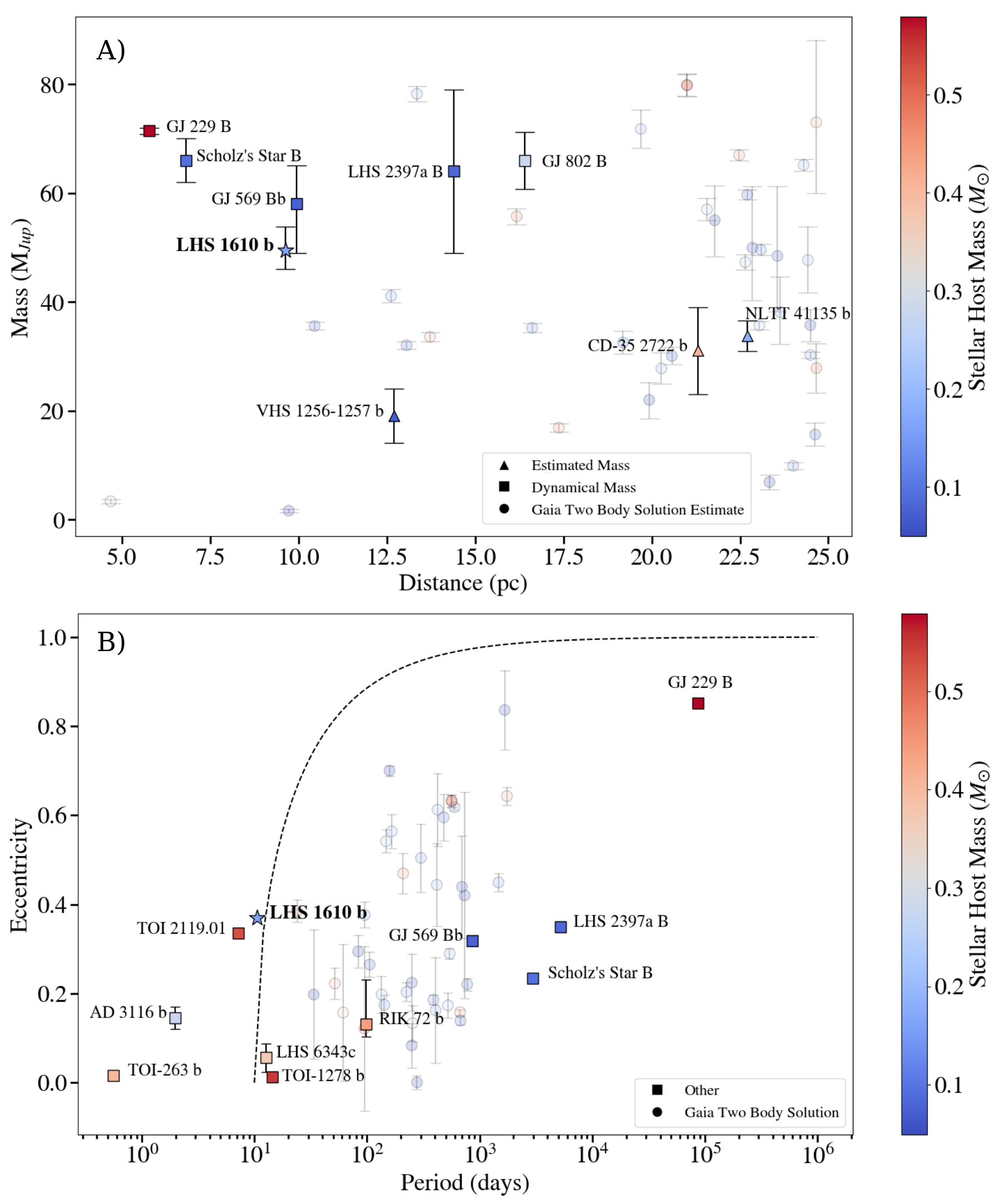}
\vspace{-0.5cm}
\end{center}
\caption{LHS~1610~b in context with other brown dwarfs orbiting M stars (BD-M systems). The colored points are confirmed BD-M star systems drawn from the NASA Exoplanet Archive, the Exoplanet.EU catalog, and the literature. Not all systems have measurements for the plot parameters, so the points on each plot are individually labeled. To highlight the yield that Gaia will help enable, the fainter points show candidate BD-M systems from Gaia two-body solutions within 25pc. Some errors are smaller than the marker size. \textbf{Top (A):} Brown dwarf masses (better than $3\sigma$) as a function of distance from Earth colored by the host star mass. LHS~1610~b is the 3rd most nearby BD-M system with a precise mass measurement. \textbf{Bottom (B):} Same as above, but showing eccentricity as a function of orbital period in days. LHS~1610~b is one of the shortest period BDs orbiting an M dwarf, and is the most eccentric in our sample besides GJ 229 B. The dashed line shows the expected eccentricity at a given period for a circularization period of 10 days. Similar to panel A, the fainter points show candidate BD-M systems from Gaia two-body solutions within 25pc.}
\label{fig:mass_precision}
\end{figure*}

LHS~1610~b joins a small but growing number of nearby brown dwarfs with precisely measured dynamical masses. Figure \ref{fig:mass_precision} puts LHS~1610~b in context with other known brown dwarf M-dwarf (BD-M) systems drawn from a compilation of objects with masses between 13 and 80 Jupiter masses from the NASA Exoplanet Archive \citep{Akeson2013}, the Exoplanet.EU catalog, and the literature. 

The top panel of Figure \ref{fig:mass_precision} shows the brown dwarf mass as a function of distance of the system from Earth where the points are color coded by the host star mass. We cut out systems that have a mass precision worse than 33\%. We find LHS~1610~b to be the third closest BD-M system within our  mass precision constraint. The other two more nearby targets are: GJ 229 B \citep{brandt2021} and Scholz's Star B \citep{Dupuy2019}. \cite{brandt2021} note that GJ 229 B is in tension with evolutionary models, as an object of its mass cannot cool to its luminosity within a Hubble time, suggesting it could instead be an unresolved binary \citep{Howe2023arxiv}. 

By a statistical analysis of the brown dwarf population around FGK stars, \cite{Ma2014} suggested that the brown dwarf population can be split into two regimes: the low mass regime ($M<42.5 M_{\text{Jup}}$) with an eccentricity distribution similar to gas giant planets, and a high mass regime with an eccentricity distribution similar to binary stars. With a mass of $49.5^{+4.3}_{-3.5}$ $M_{\text{Jup}}$, LHS~1610~b is formally in the high-mass regime discussed in \cite{Ma2014}, and therefore could have formed through molecular cloud fragmentation, similar to a binary stellar companion, as opposed to forming similar to a giant planet via gravitational instability or core-accretion. However, in reality, the boundary between the two regimes is not exact, and is rather characterized by a `depletion region' between $35 - 55 M_J$ as discussed by \cite{Ma2014}, where short-period ($P<100 \unit{days}$) brown dwarfs are observed to be intrinsically rare. From its mass alone it is unclear which formation pathway LHS~1610~b would be more compatible with. Since the \cite{Ma2014} study was performed on a sample of FGK stars, it may not be directly applicable to our BD-M dwarf binary. Further, the \cite{Ma2014} study was focused on a sample of radial velocity discovered brown dwarfs, which only have minimum mass measurements ($m \sin i$), making it unclear if the `depletion region' discussed by \cite{Ma2014} persists once true mass constraints are estimated. More discoveries of brown dwarfs that both have a RV and astrometric solutions  will allow us to measure true masses for the brown dwarfs and more robustly determine if the `depletion region' persists in the brown dwarf population.

The faint points in both panels of Figure \ref{fig:mass_precision} show estimated values for the parameters of possible BD-M systems within 25 parsecs that have Gaia two-body solutions. The stellar masses are estimated by compiling the K-magnitudes and using the \citet{mann2019} mass-luminosity relation. Any object with an estimated stellar mass less than 0.65 $\rm{M_{\odot}}$ is retained. Using the estimated stellar mass for these systems, we take the period, eccentricity, photocenter semi-major axis, parallax, and cosine of the inclination from the Gaia two-body solution and estimate the RV semi-amplitude, K, using Equation \ref{eq:K_from_a0}. With the stellar mass estimate and computed K we estimate the mass of the companion under the assumption that the Gaia two body solution correctly describes a single dark companion.
These points highlight that Gaia is starting to uncover a number of additional candidate BD-M systems, allowing further insights into their occurrence rates and eccentricity distribution. However, as highlighted in this work as well as in \cite{Halbwachs2022}, \cite{Winn2022} and \cite{marcussen2023}, precise RV follow-up observations are necessary to confirm  that the two-body solutions are accurately describing the parameters of the systems, and to robustly rule out false positive scenarios

To further investigate LHS~1610~b's association with the planet or binary star formation pathways, Figure \ref{fig:mass_precision}B compares the eccentricity and period of LHS~1610~b to other brown dwarfs. We see that LHS~1610~b has the highest eccentricity for confirmed systems with periods $<10,000$ days. Because of the short-period and non-zero eccentricity, we may expect that tides will circularize LHS~1610~b's orbit \citep[e.g.,][]{Mazeh2008,Damiani2016}. The companion will be circularized if its orbital period is less than the circularization period. M dwarf binaries have an observed circularization period of $\sim$$10$ days \citep{Udry2000, Mayor2001}, while Sun-like binaries are $\sim$$10$-$12$ days \citep{Duquennoy1991, Meibom2005, Raghavan2010}. The dashed line in Figure \ref{fig:mass_precision}B indicates the maximum eccentricity a companion of a given period could have without experiencing tidal effects if the circularization period is 10 days. \citep[Equation 3 in][]{Halbwachs2005}. With a period of $P=10.6 \unit{days}$ and an eccentricity of $e=0.37$, LHS~1610~b is inconsistent with both the aforementioned circularization period trends. Instead, LHS~1610~b fits a shorter circularization period of $\sim$8 days, more in-line with the circularization periods of a few days observed for giant exoplanets \citep{Halbwachs2005,Pont2011,Bonomo2017}.

We estimate the circularization timescale to compare with our coarse estimate of the age of $7.0_{-4.7}^{+4.5} \unit{Gyr}$ from the SED analysis. We use the Sonora-Bobcat models \citep{marley2021}, to estimate the radius of the brown dwarf based on the mass and an age between 0.5 and 12 Gyrs. Using the solar metallicity models, at 0.5 Gyr we find a radius of $0.99 R_{\text{Jup}}$, and at 12 Gyr, $0.79 R_{\text{Jup}}$. We estimated the circularization timescale of the brown dwarf using Equations 1 and 2 from \citet{Jackson2008} and presented in \citet{canas2022bd} as:
\begin{equation}
\frac{1}{\tau_e} = \frac{1}{\tau_{e,*}} + \frac{1}{\tau_{e,\text{BD}}},
\end{equation}
\begin{equation}
\frac{1}{\tau_{e,*}} = a_{\text{BD}}^{-13/2} \left(\frac{171}{16}\right) \sqrt{\frac{G}{M_*}} \frac{R_*^5 M_{\text{BD}}}{Q_*},
\end{equation}
\begin{equation}
\frac{1}{\tau_{e,\text{BD}}} = a_{\text{BD}}^{-13/2} \left(\frac{63}{4}\right) \sqrt{G M_*^3} \frac{R_{\text{BD}}^5}{Q_{\text{BD}} M_{\text{BD}}},
\end{equation}
where $\tau_e$ is the circularization timescale, $\tau_{e,*}$ is the timescale contribution from the star and $\tau_{e,\text{BD}}$ is the timescale contribution from the brown dwarf. The parameters $a_{\text{BD}}$, $M_*$,  $M_{\text{BD}}$, $R_*$, $R_{\text{BD}}$, $Q_*$, $Q_{\text{BD}}$ are the semimajor axis of the brown dwarf, the stellar host mass, the brown dwarf mass, the radius of the stellar host, the brown dwarf radius, the tidal dissipation factors of the stellar host and brown dwarf, respectively. We assume a value of $Q_\star = 10^7 $ based on modeling done in \cite{Gallet2017} and $Q_{\text{BD}} = 10^5$ as inferred for Jupiter \citep{Goldreich1966,Lainey2009,Lainey2016}. We assume the tidal dissipation factors remain constant over time, but in reality this factor will change as the star and brown dwarf evolve. For the 0.5 Gyr and 12 Gyr age assumptions, we obtain circularization timescales of 240 Gyr and 720 GYr, respectively. The high circularization timescale we obtain in both cases suggests the system is not circularizing.

Additionally, we estimated the timescale for tidal synchronization using the equation from \cite{rasio1996,guillot1996},
\begin{equation}
\tau_s = Q_{\text{BD}} \left( \frac{R_{\text{BD}}^3}{G M_{\text{BD}}} \right) \omega_{\text{BD}} \left( \frac{M_{\text{BD}}}{M_\star} \right)^2 \left(\frac{a_{\text{BD}}}{R_{\text{BD}}} \right)^6,
\end{equation}
where $\omega_{\text{BD}}$ is the primordial rotation rate of the brown dwarf. We again assume a value of $Q_{\text{BD}} = 10^5$. For the primordial spin rate, Figure 13 of \citet{Tannock2021} shows measured rotation periods for 78 L-, T-, and Y-dwarfs. Across spectral type, rotation periods between 1 and 10 hours are common. First, we will assume a rotation rate ($\omega_{BD}$) of 10 hours ($1.7 \times 10^{-4}$ cycles per second), equal to that of Jupiter. For the 0.5 Gyr and 12 Gyr age assumptions, we obtain synchronization timescales of $\sim$7 Gyrs and $\sim$15 Gyrs, respectively. If we assume a rotation period of 1 hour instead ($1.7 \times 10^{-3}$ cycles per second), then our synchronization timescales for the 0.5 Gyr and 12 Gyr age assumptions are $\sim$70 Gyrs and $\sim$150 Gyrs, respectively. Although we note that the exact value is strongly dependent on the assumed primordial spin rate and tidal dissipation factor, the lengthy timescales indicated above suggest the brown dwarf has likely not become tidally locked. 

\section{Prospects for star-companion interactions} \label{sec:interactions}

Companions that closely orbit their host stars are thought to interact magnetically or \textit{sub-Alfvénically}. These interactions are expected to enhance the activity of the host star, manifesting as enhanced X-ray and flaring activity \citep{lanza2018, ilin2022}, radio emission \citep{zarka01, zarka07, Callingham2021, kavanagh23}, and chromospheric spots \citep{shkolnik2005}. Furthermore, if the companion is magnetised, it may exhibit auroral emission due to the interactions between the stellar wind of the host star and its magnetosphere \citep{zarka01}. Signatures of these interactions in close-in exoplanetary systems are often loosely referred to as star-planet interactions. However, given that the companion of LHS~1610 is a brown dwarf, the term star-companion interactions is more appropriate. 

To date, the magnitudes and scaling relationships of these interactions remain poorly constrained. Therefore, detections of objects compatible with such interactions are valuable laboratories to study such models. Due to the system's proximity, large size, and short orbital period of the companion, LHS~1610 is a promising target for searching for possible evidence of star-companion interactions. We discuss this possibility in further detail in the following subsections.


\subsection{Investigation of Potential Companion-induced Flaring}
\label{sec:flaring}
One way to look for evidence of sub-Alfv\'enic interactions is through looking for evidence of phase dependence of flaring \citep[e.g.,][]{lanza2018,ilin2022} and/or orbital phase dependent variations in stellar activity indicators \citep[e.g.,][]{shkolnik2005}, especially at orbital phases close to periastron. Examining the 6 visits of HPF observations, we see no evidence of emission in the Ca II IRT activity indicators, and detect no hints of modulation as a function of orbital phase, and conclude that the star is not chromospherically active.

\begin{figure*}[t!]
\begin{center}
\includegraphics[width=\textwidth]{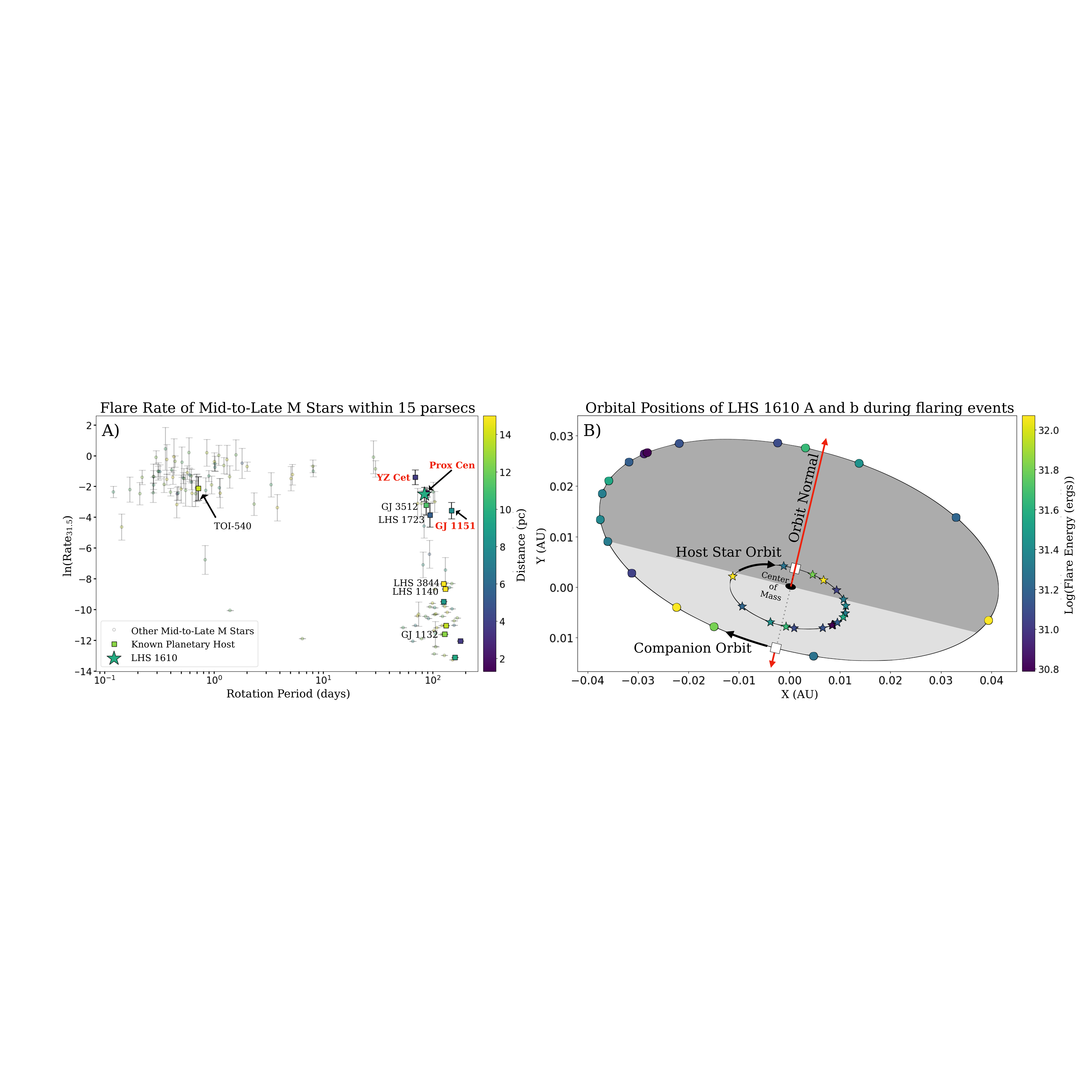}
\vspace{-0.6cm}
\end{center}
\caption{\textbf{M Dwarf Flare Rates (A):} Log flare rate of flares with energies above the $3.16 \times 10^{31} \unit{ergs}$ energy threshold discussed in \cite{Medina2022} as a function of stellar rotation period. Points are color coded by distance to Earth. LHS 1610 is highlighted with the star. Systems with substellar companions are highlighted with the black labels. Systems with published radio detections are highlighted in red. LHS 1610 joins a group of systems with high flare rates and long rotation periods. \textbf{LHS 1610 flare orbital locations (B):} Orbit visualization of the LHS~1610~b (outer ellipse) and its host star (inner ellipse) around the common center of mass using the joint Gaia-RV parameters from Table \ref{tab:results}. The orbit is inclined $\sim114^\circ$ and the orbital direction is clockwise. The red vector denotes the orbit normal, and the periastron location is highlighted with the dotted line between the white squares. Locations of flares are highlighted twice per flare with the colored points: with a star/circle marker at the corresponding star/companion location. We do not see statistically significant evidence for a phase-dependence of flare locations. The shaded regions define the intersection of the plane of the orbit with the plane of the sky along the line of nodes. A table with flare parameters (times, energies, phases) is available.}
\label{fig:flares}
\end{figure*}

To constrain the possibility of flare-induced interactions in LHS 1610, we examined the three available Sectors of TESS data of LHS 1610. To detect flares, we used the \texttt{stella} code \citep{feinstein2020stella,feinstein2020} which leverages a set of trained Convolutional Neural Networks (CNN) to identify flares. To estimate a flare rate, we follow the methods outlined in \cite{pope2021} and \cite{Medina2020}, which we summarize briefly here. Using \texttt{stella}, we analyze the TESS 2 minute cadence Presearch Data Conditioning Single-Aperture Photometry (PDCSAP) light curves using the \texttt{lightkurve} package \citep{lightkurve} and apply a flare probability threshold of 0.6. To remove false positives, we follow \cite{pope2021} and remove flares with a) fractional amplitudes less than 3 times the standard deviation of a 400-min smoothed light curve, or b) rise+fall times less than 4 minutes (two TESS cadences). After this step, sectors are reviewed by eye to add or remove any flares that were clearly misidentified.

Using this methodology, we detected a total of 17 flares: 4 in Sector 42, 6 in Sector 43, and 7 in Sector 44 (see Figure \ref{fig:flares2} in the Appendix). From this, we estimate a flare rate of $0.28 \pm 0.07$ flares per day estimated using the total number of flares across the observing baseline covered by all three sectors. We estimate the $1\sigma$ uncertainty using a two-sided Poisson confidence interval. This flare rate is high given the star's rotation period when compared to a sample of nearby M stars with confirmed radio emission presented in \cite{pope2021}. LHS~1610~b has a flare rate similar to the M stars  DO Cep and LP 259-39 which have substantially more rapid rotation periods of 0.41 days and 1.7 days, respectively. 

To further investigate the possibility of flare-induced interactions, in Figure \ref{fig:flares}A we compare the flare rate of LHS 1610 as a function of rotation period from the volume-complete sample of mid-to-late M dwarfs within 15pc from \cite{Medina2020} and \cite{Medina2022}. By replicating their energy cutoff, completeness correction, and flare energy distribution, we see that LHS 1610 has a flare rate on the high end for its stellar rotation period of $P_{\mathrm{rot}} = 84.3 \pm 8 \unit{days}$. The natural log of this flare rate is $\ln(R_{31.5}) = -2.51 \pm 0.45 \unit{flares\:day^{-1}}$ for flares above an energy of $3.16 \times 10^{31}\unit{ergs}$. We refer to this as the "high energy" flare rate. In Figure \ref{fig:flares}A, we highlight systems with confirmed substellar companions from the NASA Exoplanet Archive. Additionally, in Figure \ref{fig:flares}A we label in red systems with published radio detections that have been highlighted as potentially compatible with sub-Alfv\'enic interactions including GJ 1151 \citep{vedantham2020gj1151,Callingham2021}, Proxima Cen \citep{Perez-Torres2021}, and YZ Ceti \citep{pineda2023,trigilio2023}. Figure \ref{fig:flares}A shows that LHS 1610 joins those systems as an inactive, nearby M star with a longer rotation period, high optical flare rate, and a known companion. These similarities lead us to speculate that the brown dwarf may be inducing flares on LHS 1610. We evaluate the feasibility of observations that could be used to gain further insights into such interactions in the next section.

We further visualize the position of the flares in the orbit of LHS 1610 in Figure \ref{fig:flares}B. Each flare along with its flare energy are shown both on the host star orbit (star markers), as well as the position of the companion (circles). The flare energies are estimated following the same methodology as \cite{Medina2020}. From Figure \ref{fig:flares}, we do not see any clear phase-dependent preference, including no clear preference for flaring close to periastron. In the half of the orbit encompassing periastron (phase values between -0.25 and 0.25 in Panel D of Figure \ref{fig:flares2}) we find 9 flares, which is consistent with the expectation of $8.6 \pm 2.2$ flares given our flare rate, suggesting there is no preference for flaring near periastron. We note that due to the low number of 17 flares detected, the Poisson counting uncertainties on expected numbers of flares remain high. As such, additional flare monitoring to increase the total number of flare detections would be needed to provide conclusive evidence of any flare dependence in the system.

Lastly, brown dwarfs are known to have flares at comparable strengths to those observed around M stars \citep{Gizis2017,Paudel2020}, and some or all of the flares seen in the TESS photometry could be attributed to the brown dwarf. However, flaring brown dwarfs are generally young ($<1$ Gyr), and have spectral types earlier than L5 (${T}_{\text{eff,L}5} \sim 1800$ K). Using the Sonora-Bobcat models, we estimate the temperature of LHS~1610~b at the lower bound of our age estimate (2.5 Gyr)  and find a value of $\sim 1100$ K. Paired with the age estimate, LHS~1610~b seems unlikely to be the origin of the flares.

\subsection{Stellar Wind Environment} \label{sec:spi}
For sub-Alfv\'enic interactions to occur, the companion must orbit sub-Alfvénically, i.e., with an Alfvénic Mach number ($M_A$) less than unity \citep{saur2013}:
\begin{equation}
M_A = \Delta u / u_A < 1 .
\label{eq:mach number}
\end{equation}
Here $\Delta u$ is the relative velocity of the stellar wind as seen by the orbiting planet:

\begin{equation}
\Delta \vec{u} = \vec{u}_\mathrm{w} - \vec{u}_\mathrm{orb} ,
\end{equation}
where $u_\mathrm{w}$ is the wind velocity, and $u_\mathrm{orb}$ is the orbital velocity of the brown dwarf. $u_A$ is the Alfvén speed:
\begin{equation}
u_A = \frac{B_\mathrm{w}}{\sqrt{4\pi\rho_\mathrm{w}}},
\label{eq:alfven velocity}
\end{equation}
where $B_{\mathrm{w}}$ and $\rho_{\mathrm{w}}$ are the magnetic field strength and mass density of the stellar wind respectively. Therefore, knowing how both of these quantities varies as a function of distance, along with the wind velocity, is required to determine if a companion orbits sub-Alfv\'enically.

The wind densities of low-mass main sequence-stars like LHS~1610 are notoriously difficult to detect given their rarefied nature, and current measurements are both indirect and few in numbers. A way to quantify stellar wind densities is through their mass-loss rate:
\begin{equation}
\mdot = 4\pi r^2 \rho_\mathrm{w} u_\mathrm{w,r},
\label{eq:mdot}
\end{equation}
where $r$ the is distance from the star, and $u_\mathrm{w,r}$ is the wind velocity in the radial direction. \citet{wood21} presented the most up to date list of mass-loss rate estimates for low-mass main sequence stars. For M~dwarfs, these values vary from $0.06$ to $200~\mdot_\sun$, where $\mdot_\sun = 2\times10^{-14}~M_\sun~\textrm{yr}^{-1}$ is the mass-loss rate of the Sun \citep{cohen11}. In the absence of a measured mass-loss rate for LHS~1610, we adopt this range of values.

Stellar winds from low-mass main-sequence stars are thought to be predominantly driven by the thermal expansion of a hot corona \citep{gombosi18}. Therefore, estimating the temperature of their coronae can provide constraints on the velocity of the wind. \citet{johnstone15} found the following empirical relation between the observed surface X-ray flux ($F_X$) and inferred coronal temperatures ($T_\mathrm{corona}$) of these stars: 
\begin{equation}
T_\mathrm{corona} = 0.11{F_X}^{0.26}\times10^6 ~\unit{K},
\label{eq:coronal temp}
\end{equation}
where $F_X$ is in $10^6~\mathrm{erg~s}^{-1}~\mathrm{cm}^{-2}$. Note that the errors in the fit coefficients in Equation~\ref{eq:coronal temp} are less than 1\%. \citet{magaudda20} reported an X-ray luminosity for LHS~1610 of $L_X = 10^{26.94\pm0.04}$~erg~s$^{-1}$ from observations with the \textit{Chandra X-ray Observatory} \citep{ChandraXRT, Wright2016, Wright2018}. Converting this value to the surface X-ray flux ($F_X = L_X / 4\pi{R_\star}^2$), we estimate a coronal temperature for LHS~1610 via Equation~\ref{eq:coronal temp} of $(3.05\pm0.07)\times10^6$~K.

\begin{figure}
\centering
\includegraphics[width = \columnwidth]{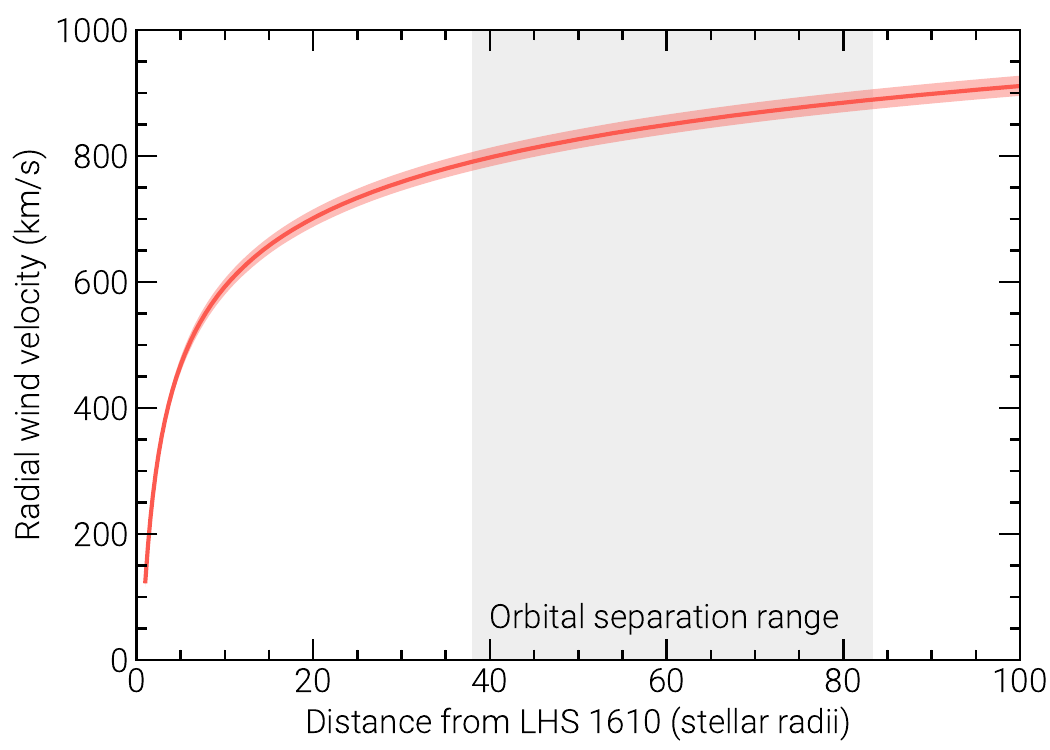}
\caption{The radial wind velocity profile estimated for LHS~1610. The red shaded region shows the $1\sigma$ uncertainty. The range of orbital separations between the brown dwarf and host star is shown by the grey shaded region. During its orbit, the brown dwarf is subjected to stellar wind velocities of $\sim800$ to $900$~km~s$^{-1}$.}
\label{fig:wind}
\end{figure}

Models for thermally-driven stellar winds have existed for decades, the first of which being proposed by \citet{parker58}. Despite its simplicity, it reproduces the bulk properties of the solar wind remarkably well. While more sophisticated models exist, they are both computationally expensive and dependent on information that is often not readily-available, such as the surface magnetic field topology \citep[e.g.,][]{kavanagh22}. Again in the absence of such information for LHS~1610, we opt for a simple prescription to estimate its wind velocity profile. For this, we use the code developed by \citet{kavanagh20}, which produces the radial wind velocity profile $u_\mathrm{w,r}$ described by \citet{parker58} for a given stellar mass and coronal temperature. We show our estimated velocity profile in Figure~\ref{fig:wind}. We find that over the course of its orbit, the companion is subject to wind speeds of around 800 to 900 km/s.

The wind velocity solution described by \citet{parker58} is purely radial in direction. Such a scenario is expected for a star like LHS~1610 given its slow rotation \citep{preusse05, johnstone17}. Similarly, it is expected to have a large-scale magnetic field that is predominantly radial. To validate this, we take the wind model for the star Proxima Centauri presented by \citet{kavanagh21}, and compute the radial component of the wind velocity and magnetic field, as well as the fraction of the large-scale magnetic field that is open, as a function of distance. This is shown in Appendix~\ref{sec:radial field}. We choose Proxima Centauri given both its similarities to LHS~1610 (Section~\ref{sec:flaring}), and also since we lack a mass-loss rate estimate and the magnetic field topology of the star (unlike Proxima). We find that the wind velocity and magnetic field are predominantly radial at a distance greater than around 10 stellar radii for Proxima Centauri. This is also consistent with an open magnetic field geometry.

Knowing the geometry of the large-scale magnetic field is important for estimating signatures of possible star-companion interactions, as we will see in the following sections. For an open-field geometry, magnetic flux conservation gives us the following scaling for the magnetic field of the wind:
\begin{equation}
B_\mathrm{w} = \Bv \Big(\frac{R_\star}{r}\Big)^2 .
\end{equation}
Here $\Bv$ is the unsigned average large-scale magnetic field strength at the stellar surface. This is the most relevant quantity in terms of what drives the stellar wind outflow \citep{jardine17, vidotto21}. 

The Zeeman Doppler imaging method (ZDI) can provide estimates for $\Bv$. Recently, \citet{klein21a}, \citet{bellotti23b}, and \citet{lehmann24} used ZDI to estimate magnetic field strengths for slowly-rotating M~dwarfs like LHS~1610 for the first time. These stars, all with rotation periods exceeding 40 days, exhibit predominantly dipolar magnetic fields, with strengths ranging from $\Bv = 16$ to $214$~G. We adopt this range of values for LHS~1610. We note that \citet{henry18} illustrated that LHS~1610 may be undergoing magnetic cycles. However, the values used here from \citet{lehmann24} include those for stars exhibiting magnetic variability and long-term activity cycles. Therefore, any variability in the large-scale surface magnetic field of LHS~1610 should be encoded within the range of values adopted here.

With the wind velocity profile, mass-loss rate, and magnetic field of LHS~1610 estimated, we now compute the Alfv\'en velocity via Equation~\ref{eq:alfven velocity}. The density profile is obtained from Equation~\ref{eq:mdot}. Note that since the orbital velocity vector of the brown dwarf is perpendicular to the radial direction, and the wind velocity is in the radial direction, the relative velocity between the wind and the orbit of the brown dwarf is $\Delta u = \sqrt{{u_\mathrm{w,r}}^2 + {u_\mathrm{BD}}^2}$. With this, we then compute the Alfv\'enic Mach number via Equation~\ref{eq:mach number}. 

\begin{figure}
\centering
\includegraphics[width = \columnwidth]{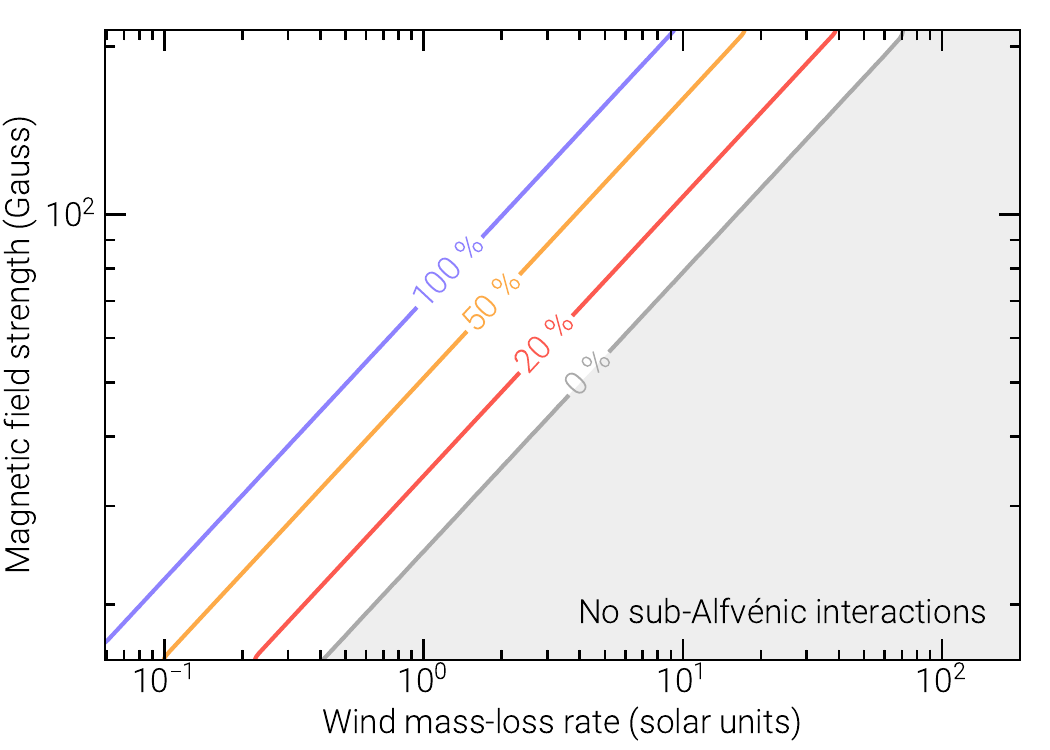}
\caption{The fraction of the orbit that LHS~1610~b orbits sub-Alfvénically as a function of the stellar wind mass-loss rate and unsigned average large-scale surface magnetic field of the host star. The grey shaded region shows where sub-Alfvénic interactions are not possible. Provided the mass-loss rate is sufficiently low and the surface magnetic field is sufficiently strong, the brown dwarf can magnetically interact with the host star, enhancing the star's activity.}
\label{fig:sub alfvenic}
\end{figure}

Given that the separation between the star and its companion varies significantly over the orbit, we compute the fraction of the orbit where the companion is sub-Alfv\'enic as a function of the mass-loss rate and magnetic field strength at the surface. This is shown in Figure~\ref{fig:sub alfvenic}. We see that sub-Alfv\'enic interactions are possible, provided the wind mass-loss rate is low enough and the surface magnetic field is strong enough. However, currently we do not have sufficient statistics for M~dwarfs to estimate the probability density of the mass-loss rate and surface field strength for LHS~1610, and therefore cannot determine the likelihood that it has conditions sufficient to drive sub-Alfv\'enic interactions. The necessity of  the surface magnetic field measurement to study star-companion interations motivates follow-up observations to derive the surface magnetic field of the star via the ZDI method.

\subsection{Sub-Alfv\'enic interactions between LHS~1610~b and its host star}\label{sec:sub alf}

If LHS~1610~b is in a sub-Alfvénic orbit, it could enhance emission on the host star at a wide range of wavelengths by the dissipation of energy carried by Alfvén waves generated via sub-Alfvénic interactions \citep{zarka07, saur2013}. The power produced via these interactions is \citep{saur2013, kavanagh22}:
\begin{equation}
P_\mathrm{SA} = \pi^{1/2} {R_\mathrm{obs}}^2 B_\mathrm{w} {\rho_\mathrm{w}}^{1/2} \Delta u^2 \sin^2\theta ,
\label{eq:power sub alfvenic}
\end{equation}
where $R_\mathrm{obs}$ is the size the obstacle perturbing the stellar magnetic field, and $\theta$ is the angle between the relative velocity and magnetic field vectors $\Delta\vec{u}$ and $\vec{B}_\mathrm{w}$. As shown in Section \ref{sec:spi}, the magnetic field of the wind likely points in the radial direction (i.e. $\theta$ is also the angle between $\Delta\vec{u}$ and the wind velocity vector $\vec{u}_\mathrm{w}$). In the case that LHS~1610~b is unmagnetised, $R_\mathrm{obs}$ is simply its radius $R_\mathrm{BD}$. However, if it possesses an intrinsic magnetic field, the size of the obstacle is the size of its magnetosphere $R_\mathrm{M}$. This can be estimated by computing the magnetopause distance, the point where the pressure of the incident stellar wind balances with the pressure exerted by the magnetic field of LHS~1610~b \citep{vidotto12}: 
\begin{equation}
R_\mathrm{M} = \Big(\frac{{B_\mathrm{BD}}^2}{32\pi p_\mathrm{w}}\Big)^{1/6}~R_\mathrm{BD} .
\label{eq:magnetosphere size}
\end{equation}
Here $B_\mathrm{BD}$ is the field strength of LHS~1610~b at its magnetic poles, assuming the field is dipolar, and $p_\mathrm{w}$ is the pressure of the stellar wind at its orbit:
\begin{equation}
p_\mathrm{w} = \rho_\mathrm{w}(a^2 + \Delta u^2) + \frac{{B_\mathrm{w}}^2}{8\pi} ,
\end{equation}
where $a = \sqrt{2kT/m_p}$ is the isothermal sound speed.

As there are no published detections of radio aurorae from LHS~1610~b, we cannot estimate its magnetic field strength directly \citep{kao18}. Therefore, we use the following theoretical prescription for the magnetic field strength from \citet{reiners10} to estimate the dipolar magnetic field strength of LHS~1610~b:
\begin{equation}
B_\mathrm{BD} = 3.39 \Big(\frac{M_\mathrm{BD} {L_\mathrm{BD}}^2}{{R_\mathrm{BD}}^7}\Big)^{1/6}~\mathrm{kG},
\label{eq:bd field strength}
\end{equation}
where $M_\mathrm{BD}$, $R_\mathrm{BD}$, and $L_\mathrm{BD}$ are the mass, radius, and luminosity of the brown dwarf. Note that \updated{the values are in solar units in Equation \ref{eq:bd field strength} and that} this relation assumes that the brown dwarf is rotating sufficiently fast which is likely for LHS~1610~b given its long synchronization time (Section~\ref{sec:comparison}). 

The radius and luminosity of brown dwarfs can be estimated from isochrones if its mass and age are known. However, the uncertainty on the age of the system is large. Therefore, we consider three ages of 2, 7, and 12 Gyrs for LHS~1610~b. We draw samples for the mass of the brown dwarf from the constraint obtained in Section~\ref{sec:adopted parameters}, and linearly interpolate for the radius and luminosity from the the solar metallicity isochrones provided by \citet{marley2021} at the three ages listed above. In this age range, we find radii of 0.86 to 0.79~$\Rjup$, and luminosities of $-4.88$ to $-5.93~\Lsun$ in log-space, where $\Lsun$ is the luminosity of the Sun \citep[$\Lsun = 3.83\times10^{33}$~erg~s$^{-1}$,][]{vieira22}. 

With the radius and luminosity estimated, we now use Equation~\ref{eq:bd field strength} to estimate the field strength for LHS~1610~b. For this, we find strengths of 816 to 402~G, which are up to an order of magnitude larger than the field strengths estimated for M~dwarfs similar to LHS~1610. This is unsurprising however, as the star likely rotates much slower than the brown dwarf. Having estimated the radius and field strength of the brown dwarf, we finally can estimate the size of its magnetosphere via Equation~\ref{eq:magnetosphere size} in combination with the estimated stellar wind properties. We find large sizes, ranging from around 7 to 13 Jupiter radii depending on the age and orbital phase. The inferred parameters and their uncertainties as a function of age are listed in Table~\ref{table:BD values}.

\begin{table*}[]
\caption{Physical parameters estimated for LHS~1610~b as a function of age. The radius $R_\mathrm{BD}$ and luminosity $L_\mathrm{BD}$ are obtained via our estimate for the mass of $49.5^{+4.3}_{-3.5}~$ Jupiter masses using the isochrones from \citet{marley2021}. We then use these to compute the dipole strength $B_\mathrm{BD}$ via Equation~\ref{eq:bd field strength}. Combined with the estimated stellar wind parameters, we estimate the range of sizes for the brown dwarf's magnetosphere $R_\mathrm{M}$, which varies over its orbit due to the eccentricity. The lower and upper limits corresponds to the values computed at periastron and apastron respectively.}
\centering
\begin{tabular}{ccccc}
\hline
Age & $R_\mathrm{BD}$ & $\log L_\mathrm{BD}$ & $B_\mathrm{BD}$ & $R_\mathrm{M}$ \\
(Gyr) & ($\Rjup$) & ($\Lsun$) & (G) & ($\Rjup$) \\
\hline
2 & $0.86\pm0.01$ & $-4.88^{+0.12}_{-0.10}$ & $816^{+98}_{-78}$ & $9.67^{+1.67}_{-0.78}$ – $12.50^{+2.50}_{-1.06}$ \\
7 & $0.80\pm0.01$ & $-5.63^{+0.11}_{-0.09}$ & $495^{+61}_{-46}$ & $7.67^{+1.34}_{-0.61}$ – $9.91^{+2.00}_{-0.83}$ \\
12 & $0.79\pm0.01$ & $-5.93^{+0.11}_{-0.08}$ & $402^{+48}_{-35}$ & $7.02^{+1.23}_{-0.56}$ – $9.07^{+1.83}_{-0.76}$\\
\end{tabular}
\label{table:BD values}
\end{table*}

With both the estimates for the properties of the stellar wind and the obstacle size of the brown dwarf established, we now use Equation~\ref{eq:power sub alfvenic} to estimate the power dissipated along the stellar magnetic field via sub-Alfv\'enic interactions. We uniformly sample the mass-loss rate and unsigned large-scale surface magnetic field strength in the aforementioned ranges adopted for LHS~1610. We also consider the cases where LHS~1610~b is both magnetised and weakly/unmagnetised (i.e., $R_\mathrm{obs} \approx R_\mathrm{BD}$ in Equation~\ref{eq:power sub alfvenic}). 

\begin{figure}[t!]
\centering
\includegraphics[width = \columnwidth]{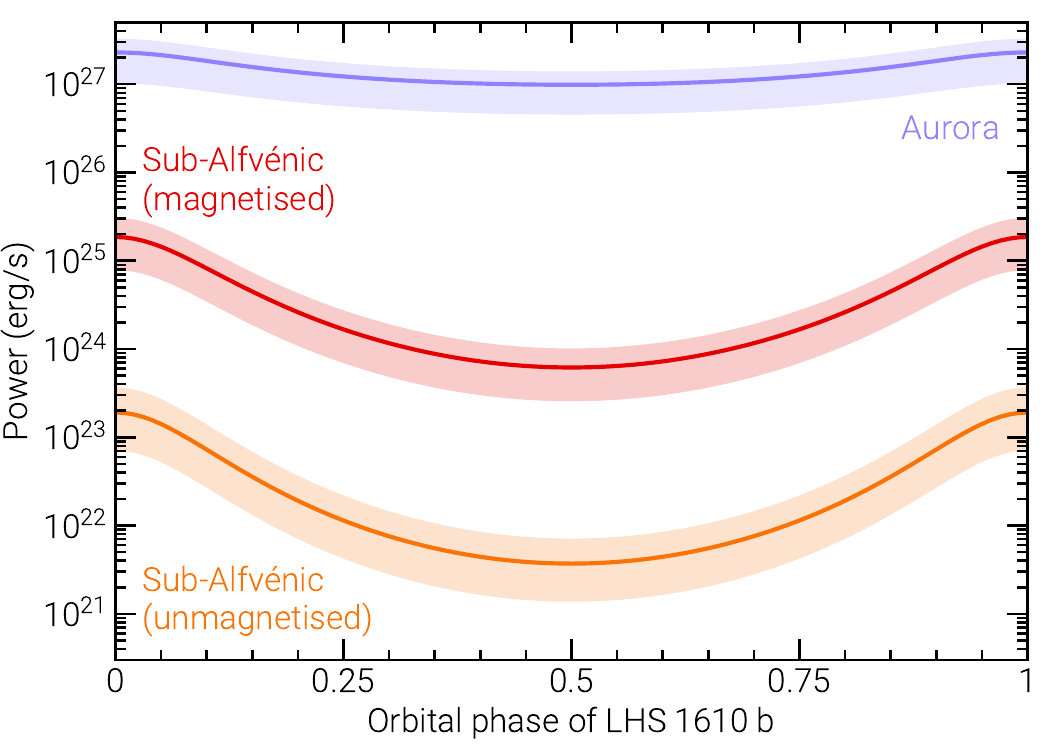}
\caption{Estimated powers from both sub-Alfv\'enic interactions between LHS~1610~b and its host star (red and orange curves), and auroral emission from the brown dwarf's magnetosphere due to the incident stellar wind (purple curves). Each solid line and shaded region shows the median and the $1\sigma$ region, respectively. The powers are shown as a function of orbital phase, which are modulated due to the eccentricity of the orbit. For the sub-Alfv\'enic powers, we show the cases where the brown dwarf is both magnetised (red curves) and unmagnetised (orange curves). The powers are computed assuming an age of 7~Gyr for the brown dwarf. The differences in the curves at the ages of 2 and 12 Gyr are small, as the radius of the brown dwarf does not change significantly, and the size of its magnetosphere varies weakly with the magnetic field strength (see Equation~\ref{eq:magnetosphere size}). In all cases, a fraction of the energy ($10^{-4}-10^{-2}$) is expected to be released at radio wavelengths, resulting in flux estimates that are within reach of sensitive radio telescopes such as LOFAR and the VLA. Note that the sub-Alfv\'enic interaction scenario is only possible provided the host star has a sufficiently low mass-loss rate and strong surface magnetic field (see Figure~\ref{fig:sub alfvenic}).}
\label{fig:spi power}
\end{figure}

In Figure~\ref{fig:spi power} we show the estimated power generated via sub-Alfv\'enic interactions between the brown dwarf and stellar magnetic field as a function of orbital phase, for an age of 7~Gyr. For the magnetised case, we estimate the total power produced to be within the range of $\sim10^{24}$ to $10^{25}~\textrm{erg~s}^{-1}$, whereas in the unmagnetised case, it ranges from $\sim10^{22}$ to $10^{23}~\textrm{erg~s}^{-1}$. Due to the eccentricity of the orbit, the distance between the two bodies varies from 38 to $83~R_\star$ over the course of the orbit. As a result, the brown dwarf is subjected to time-varying stellar wind conditions. Therefore, the power dissipated onto the star also varies, as is seen in the lightcurves in Figure \ref{fig:spi power}.

If the estimated power generated by sub-Alfv\'enic interactions $P_\mathrm{SA}$ is indeed dissipated onto the star's magnetic field, some fraction is expected to manifest as bright radio emission via the electron cyclotron maser (ECM) instability \citep{zarka07, saur2013}. For the Io-induced emission on Jupiter, this fraction or `efficiency ratio' $\eta$ is estimated to be around $10^{-4}$ to $10^{-2}$ \citep{turnpenney18, saur21}. If we assume this fraction of the total power is uniformly dissipated in the radio over the frequency range $\Delta\nu$, the flux density observed at a distance $d$ is:
\begin{equation}
F_\nu = \frac{\eta P_\mathrm{SA}}{d^2\Omega_\mathrm{beam}\Delta\nu} ,
\label{eq:radio flux density}
\end{equation}
where $\Omega_\mathrm{beam}$ is the solid angle of the emission beam, which is a hollow cone since ECM emission is beamed. Typical values assumed for $\Omega$ range from 0.16 to 1.6 steradian \citep{zarka04, turnpenney18}. If the frequency range $\Delta\nu$ is sufficiently wide, it is effectively equal to the cutoff frequency $\nu_\mathrm{max}$, which we assume to be that at the stellar surface. Given that ECM emission occurs at the local cyclotron frequency \citep{dulk85}, the cutoff frequency is therefore the cyclotron frequency at the surface:
\begin{equation}
\nu_\mathrm{max} = 2.8 \Bv~\textrm{MHz} ,
\end{equation}
where $\Bv$ is the unsigned surface average field strength in Gauss.

Let's consider a conservative scenario, in which $\eta = 10^{-6}$ and $\Omega = 1.6$~sr. For the unmagnetised and magnetised cases, the sub-Alfv\'enic radio fluxes produced can reach up to 0.6~mJy and 60 mJy respectively. The emission frequency varies from 45 to 600~MHz based on the estimated surface field strength. Emission in this range at a mJy level is well within the capabilities for detection with radio telescopes such as LOFAR and GMRT \citep[see][]{narang21, Callingham2021}. Therefore, LHS 1610 is a compelling case for follow-up searches in the radio for signatures of sub-Alfv\'enic interactions. 

We note also that since ECM emission is beamed, the visibility of the radio emission is highly dependent on the underlying system geometry \citep[e.g.,][]{kavanagh22, kavanagh23}. By combining the orbital characteristics derived in this work with efforts to map the surface magnetic field topology via Zeeman Doppler imaging \citep[e.g.,][]{klein21a}, we could estimate at what orbital phases we would expect to see the radio emission. In addition to potential radio emission, signatures of sub-Alfv\'enic interactions occurring in the system may be visible at different wavelengths such as in the optical (see Section \ref{sec:flaring}), UV, and infrared \citep{shkolnik2005, klein22}.

\subsection{Auroral emission from LHS~1610~b}\label{sec:auroral}
Regardless of whether LHS~1610~b is in a sub-Alfv\'enic orbit or not, if it possesses an intrinsic magnetic field, it may also exhibit strong auroral emission produced via the dissipation of the energy carried by the stellar wind onto its magnetosphere \citep{zarka07}. For the magnetised bodies in the solar system, their observed auroral emission at radio wavelengths scales linearly with the solar wind power intercepted by their magnetospheres. The power intercepted is \citep{zarka07}:
\begin{equation}
P_\mathrm{aurora} = \varepsilon \Delta u \pi {R_\mathrm{M}}^2,
\label{eq:power aurora}
\end{equation}
where $\varepsilon = \varepsilon_\mathrm{kin} + \varepsilon_\mathrm{mag} + \varepsilon_\mathrm{th}$ is the energy density of the incident wind, which is comprised of a kinetic, magnetic, and thermal component:
\begin{equation}
\varepsilon_\mathrm{kin} = \rho_\mathrm{w} \Delta u^2,
\end{equation}
\begin{equation}
\varepsilon_\mathrm{mag} = \frac{{B_\mathrm{w}^2\sin^2\theta}}{4\pi},
\end{equation}
\begin{equation}
\varepsilon_\mathrm{th} = a^2 \rho.
\end{equation}

In the solar system, generally only the kinetic and thermal components are considered, which both appear to produce powers that are directly proportional to the observed auroral radio power, with around $10^{-5}$ to $10^{-3}$ of the incident power being dissipated \citep{zarka07}. For completeness, we also include the thermal energy incident on the magnetosphere \citep{elekes23}. In a similar manner to Section~\ref{sec:sub alf}, we compute the incident power on the magnetosphere of LHS~1610~b as a function of its orbital phase, varying $\mdot$ and $\Bv$ over the assumed ranges for the host star. This is shown in Figure~\ref{fig:spi power} alongside the powers estimated in Section~\ref{sec:sub alf} from sub-Alfv\'enic interactions. We again find large powers of around $\sim10^{27}~\textrm{erg~s}^{-1}$. Note that accounting for the surface magnetic field of the host star will introduce further modulation to the powers shown in Figure~\ref{fig:spi power} \citep[e.g.,][]{kavanagh22}. Additionally, we assume the rotation and magnetic axes of the brown dwarf are aligned with the star's rotation axis. Varying the orientation of these axes could also modulate power for these interactions, in that it will alter how the magnetic fields of the wind and the brown dwarf interact. Accounting for this however would require a full 3D magnetohydrodynamics simulation, which is beyond the scope of this work.

We now compute the radio flux via Equation~\ref{eq:radio flux density}, replacing $P_\mathrm{SA}$ with $P_\mathrm{aurora}$. We again choose conservative values for $\eta = 10^{-6}$ (an order of magnitude lower than the minimum value estimated for auroral radio emission in the solar system) and $\Omega = 1.6$~sr. We set the cyclotron frequency at the magnetic poles of the brown dwarf as the cutoff frequency, as the field lines on which the aurora is driven likely connect back near the magnetic poles due to the large size of its magnetosphere. At an age of 7~Gyr, this frequency is $\sim1.4$~GHz. We find that with these estimates, the flux density should be of the order of $\sim200$~mJy. At GHz frequencies, the VLA would be suitable for follow-up observations \citep[e.g.,][]{villadsen19}, which to date has been the primary telescope used to discover radio emission from brown dwarfs \citep{tang22}. This highlights the benefit of carrying out a multiwavelength radio campaign of the system, in that we could determine if sub-Alfv\'enic interactions and/or aurorae could be detected by observing at both MHz and GHz frequencies. Such observations could constrain both the sub/super-Alfv\'enic nature of the companion, and also the field strengths of both objects. Again, the visibility of this emission is dependent on the geometry of the brown dwarf's magnetic field, which is unknown. However, the auroral signature should appear at specific rotation phases \citep{pineda17}, unlike stochastic flares from the star which should have no preferential phase. Auroral emission is also a possibility at other wavelengths. If the brown dwarf was isolated, prospects for detecting its aurora would likely be most favourable at UV wavelengths \citep{saur21}.

\section{Summary} \label{sec:summary}
We studied the LHS 1610 system, a nearby M5 dwarf ($d=9.7 \unit{pc}$) hosting a brown dwarf in a short period $P=10.6 \unit{day}$ eccentric ($e=0.37$) orbit. This system is the second closest M dwarf with a substellar companion and a Gaia two-body solution behind GJ 876, an M dwarf system hosting at least four known planets, and LHS~1610~b is the most eccentric brown dwarf orbiting an M dwarf behind GJ 229 B.

Jointly modeling the available RVs from HPF and TRES with the Gaia two-body solution, we are able to make new estimates for all of the orbital elements of LHS~1610~b. We obtain an orbital inclination of \resinc, resulting in a mass estimate of \resm, confirming the brown dwarf nature of the companion. We highlight the discrepancy between the RV-only fit eccentricity ($e = 0.37$) and that of the Gaia two-body solution ($e = 0.52$). To account for this discrepancy, we include an uncertainty scaling factor in the astrometric covariance matrix that inflates the overall errors. We note the necessity to revisit this system when the astrometric data are released in Gaia DR4.

Due to LHS~1610~b's large radius and close-in orbit around its host star, LHS~1610~b is a promising target for the detection of potential sub-Alfvénic interactions at a wide range of wavelengths. Using the available TESS data, we derive a flare rate of $0.28 \pm 0.7 \unit{flares/day}$. When accounting for a flare energy cut-off ($E>10^{31.5}$), the subsequent flare rate places LHS 1610 among the high end for its rotation period amongst a volume-complete sample of mid-to-late M stars within 15pc from \cite{Medina2020} and \cite{Medina2022}. Within this sample, LHS 1610 is similar in spectral type, flare rate, and rotation period to Proxima Centauri and YZ Ceti, both of which have observed radio emission attributed to possible sub-Alfv\'enic interactions. Using the available TESS data for LHS 1610, we detected no significant phase-dependence of the flares, and highlight that additional data would be needed to decisively confirm or rule out such a dependence.

We simulated the expected energetics of both sub-Alfv\'enic interactions and auroral emission from the brown dwarf. For the sub-Alfv\'enic interactions, we demonstrate that LHS~1610~b may reside in a sub-Alfvénic orbit over its variable range of orbital distances due to its eccentricity, a necessary requirement to support sub-Alfv\'enic interactions. We show that given even conservative estimates, the radio emission expected from the star due to these interactions is within the sensitivity range of LOFAR and other radio instruments. Additionally, we show that direct auroral emission from the brown dwarf could be even more easily detectable than the radio emission from sub-Alfv\'enic interactions, with nominal expected radio powers in in the $10^{24} - 10^{26} \unit{erg/s}$ range. The detection of either of these interactions is dependent on the orbital phase of the brown dwarf, which benefit from the full orbital solution and precise ephemeris provided in Table \ref{tab:results}.

This work provides an outline for leveraging a fully characterized orbital solution to study star-planet/star-companion interactions. This will be useful as more systems like LHS 1610 are discovered and characterized jointly by Gaia astrometry and precise ground based radial velocities.

\facilities{HPF/HET 10m, TRES, TESS, \textit{Gaia}.} 
\software{\texttt{astropy} \citep{astropy2013},
\texttt{barycorrpy} \citep{kanodia2018}, 
\texttt{corner.py} \citep{dfm2016}, 
\texttt{emcee} \citep{dfm2013},
\texttt{Jupyter} \citep{jupyter2016},
\texttt{matplotlib} \citep{hunter2007},
\texttt{numpy} \citep{vanderwalt2011},
\texttt{pandas} \citep{pandas2010},
\texttt{pyde} \citep{pyde},
\texttt{radvel} \citep{fulton2018},
\texttt{SERVAL} \citep{zechmeister2018},
\texttt{HxRGproc} \citep{Ninan2018}.}

\newpage


\vspace{1cm}

\textbf{Acknowledgements:}
We thank the anonymous referee for the thoughtful comments and suggestions that improved the quality of the manuscript. GS acknowledges support provided by NASA through the NASA Hubble Fellowship grant HST-HF2-51519.001-A awarded by the Space Telescope Science Institute, which is operated by the Association of Universities for Research in Astronomy, Inc., for NASA, under contract NAS5-26555. RDK acknowledges funding from the Dutch Research Council (NWO) for the e-MAPS (exploring magnetism on the planetary scale) project (project number VI.Vidi.203.093) under the NWO talent scheme Vidi.

This work was partially supported by funding from the Center for Exoplanets and Habitable Worlds. The Center for Exoplanets and Habitable Worlds is supported by the Pennsylvania State University, the Eberly College of Science, and the Pennsylvania Space Grant Consortium. CIC acknowledges support by NASA Headquarters through an appointment to the NASA Postdoctoral Program at the Goddard Space Flight Center, administered by ORAU through a contract with NASA. This work was performed for the Jet Propulsion Laboratory, California Institute of Technology, sponsored by the United States Government under the Prime Contract 80NM0018D0004 between Caltech and NASA. We acknowledge support from NSF grant AST-1909506, AST-190950, AST-1910954, AST-1907622. Computations for this research were performed on the Pennsylvania State University’s Institute for Computational \& Data Sciences (ICDS). We acknowledge support from NASA XRP Grant 80NSSC24K0155.

These results are based on observations obtained with the Habitable-zone Planet Finder Spectrograph on the HET. We acknowledge support from NSF grants AST 1006676, AST 1126413, AST 1310875, AST 1310885, and the NASA Astrobiology Institute (NNA09DA76A) in our pursuit of precision radial velocities in the NIR. We acknowledge support from the Heising-Simons Foundation via grant 2017-0494. This research was conducted in part under NSF grants AST-2108493, AST-2108512, AST-2108569, and AST-2108801 in support of the HPF Guaranteed Time Observations survey. The Hobby-Eberly Telescope is a joint project of the University of Texas at Austin, the Pennsylvania State University, Ludwig-Maximilians-Universitat Munchen, and Georg-August Universitat Gottingen. The HET is named in honor of its principal benefactors, William P. Hobby and Robert E. Eberly. The HET collaboration acknowledges the support and resources from the Texas Advanced Computing Center. We thank the Resident astronomers and Telescope Operators at the HET for the skillful execution of our observations with HPF.

BJSP acknowledges and pays respect to the traditional owners of the land on which the University of Queensland is situated, and to their Ancestors and descendants, who continue cultural and spiritual connections to Country.

This research has made use of data obtained from or tools provided by the portal exoplanet.eu of The Extrasolar Planets Encyclopaedia.

This work has made use of data from the European Space Agency (ESA) mission {\it Gaia} (\url{https://www.cosmos.esa.int/gaia}), processed by the {\it Gaia} Data Processing and Analysis Consortium (DPAC, \url{https://www.cosmos.esa.int/web/gaia/dpac/consortium}). Funding for the DPAC has been provided by national institutions, in particular the institutions participating in the {\it Gaia} Multilateral Agreement.

This research made use of Astropy, a community-developed core Python package for Astronomy \citep{astropy2013}.

We acknowledge the Texas Advanced Computing Center (TACC) at The University of Texas at Austin for providing high performance computing, visualization, and storage resources that have contributed to the results reported within this paper.

\textit{TESS} data presented in this paper were obtained from the Mikulski Archive for Space Telescopes (MAST) at the Space Telescope Science Institute. The specific observations analyzed can be accessed via \dataset[https://doi.org/10.17909/fwdt-2x66]{https://doi.org/10.17909/fwdt-2x66} and \dataset[https://doi.org/10.17909/t9-nmc8-f686]{https://doi.org/10.17909/t9-nmc8-f686}. STScI is operated by the Association of Universities for Research in Astronomy, Inc., under NASA contract NAS5–26555. Support to MAST for these data is provided by the NASA Office of Space Science via grant NAG5–7584 and by other grants and contracts.

This work makes use of the \citet{PSCompPars} which is maintained by the NASA Exoplanet Science Institute at IPAC, which is operated by the California Institute of Technology under contract with the National Aeronautics and Space Administration.


\clearpage

\appendix
\section{TESS Flares}\label{sec:appendix_flares}
\begin{figure*}[h]
\centering
\includegraphics[width=\textwidth]{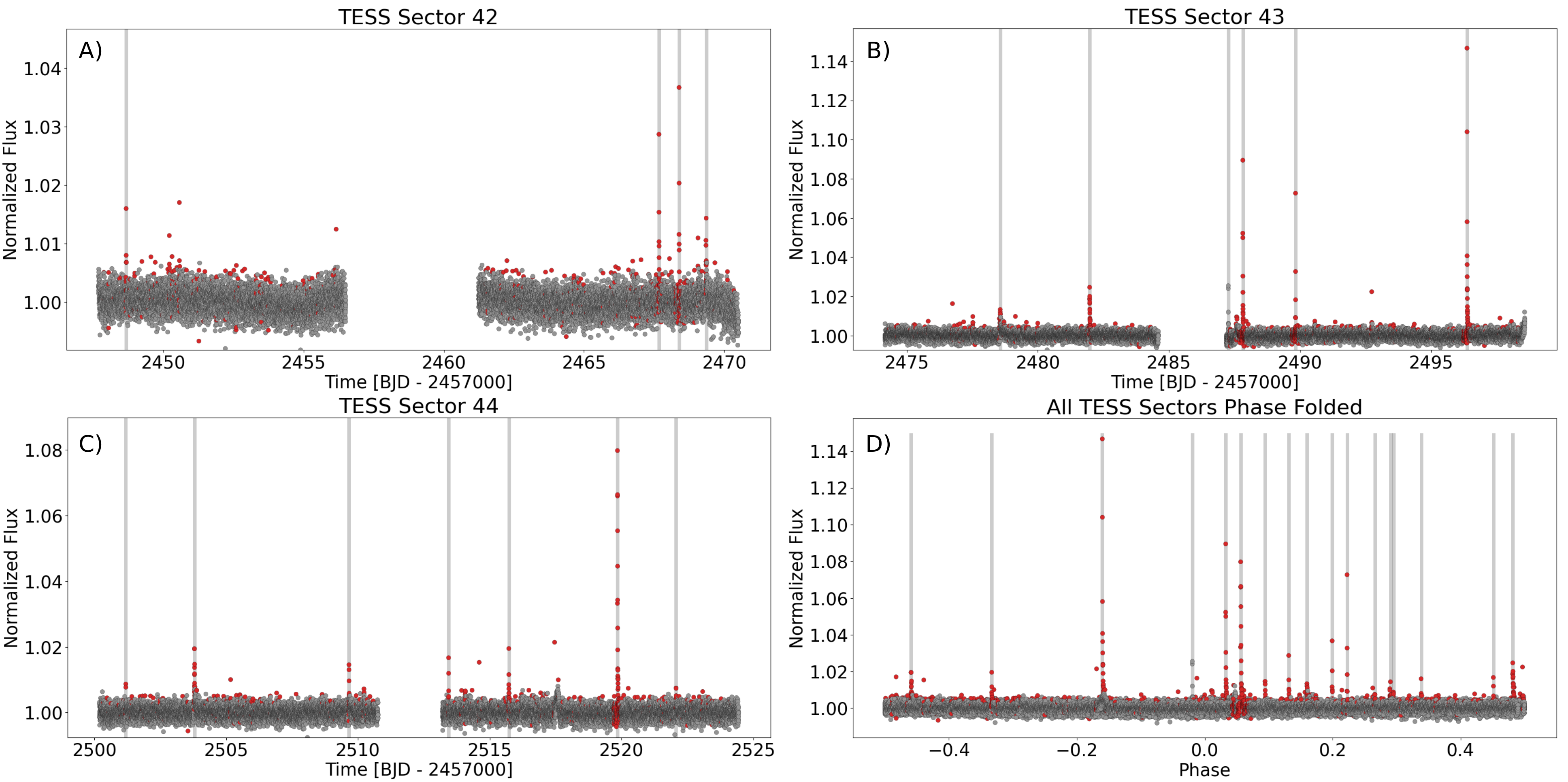}
\caption{\textbf{Flares in TESS Photometry:} Panels \textbf{A},\textbf{B}, and \textbf{C} show individual sector light curves for TESS Sectors 42, 43, and 44 respectively. Points are colored by flare probability assigned by \texttt{stella}: Red show a probability $>60\%$, and grey points less than a probability of $<60\%$. Flares that passed our cuts are highlighted by the vertical grey bars. Panel \textbf{D} contains the combination of all sectors phase folded on the period centered on the time of periastron. Flares are again highlighted by vertical grey bars. We do not see statistically significant evidence for phase-dependent flaring.}
\label{fig:flares2}
\end{figure*}

We show the TESS photometry in Figure \ref{fig:flares2} and highlight the flares identified from our analysis using the \texttt{stella} \citep{feinstein2020stella} flare-finding algorithm. The flares are highlighted by vertical grey bars. Points are color coded by the flare probability assigned by \texttt{stella}, where red points have a probability greater than 0.6, and grey points are those below 0.6. We show the individual TESS Sectors 42, 43, and 44 in panels A, B, and C, respectively. Panel D shows the combined data from all sectors, phase folded on the orbital period and centered at the periastron time. As highlighted in Section \ref{sec:flaring}, we don't see evidence for phase-dependence of the flares.

\section{Magnetic field structure surrounding LHS 1610}
\label{sec:radial field}

\begin{figure}
\centering
\includegraphics[width = 0.5\textwidth]{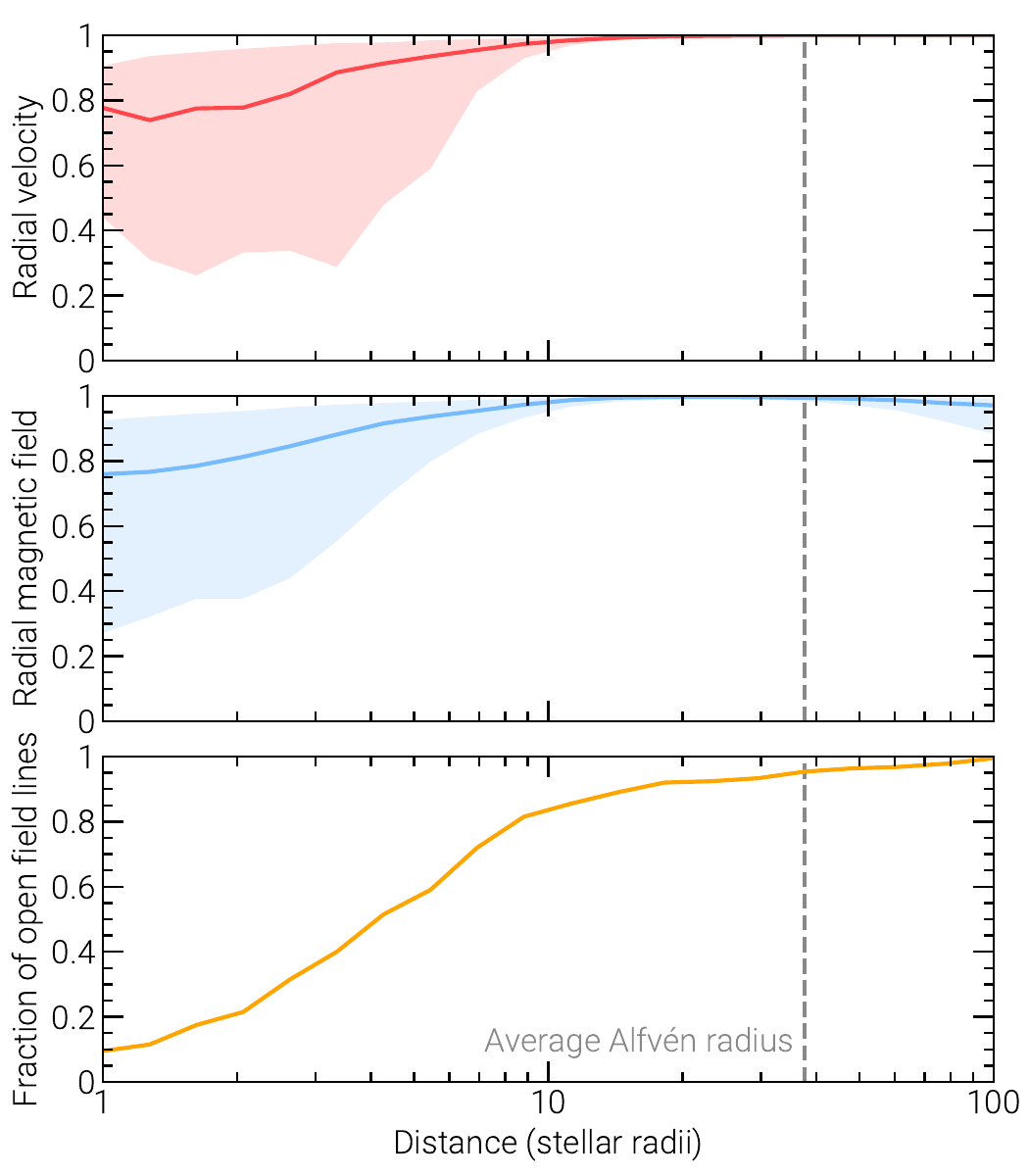}
\caption{The fraction of the wind velocity (top) and magnetic field vectors (middle) that points in the radial direction, and the fraction of open field lines (bottom) as a function of distance for the wind of the LHS 1610 analog Proxima Centauri. The shaded regions in the top two panels show the $1\sigma$ interval, highlighting the variations in the wind inside the closed field. At distances greater than around 10 stellar radii, the wind flow direction and magnetic field becomes aligned with the radial direction. Note that this happens at a distance significantly smaller than the average size of the Alfv\'en radius, inside of which sub-Alfv\'enic interactions can occur between the star and companion, enhancing the activity of the star. Therefore, most sub-Alfv\'enic orbits experience a radial magnetic field.}
\label{fig:radial field}
\end{figure}

Here, we show the alignment of the wind velocity and magnetic field vectors in the radial direction, and the fraction of open magnetic field lines, as a function of distance for the star Proxima Centauri (Figure~\ref{fig:radial field}) serving as an analog to LHS 1610. We use the stellar wind model presented in \citet{kavanagh21} for this. We see that the wind is predominantly oriented in the radial direction for distances greater than around 10 stellar radii, both in terms of the mass flow and magnetic field. This is necessary information for computing the angle $\theta$ between the vectors $\Delta\vec{u}$ and $\vec{B}_\mathrm{w}$, which in turn allows for the power dissipated via sub-Alfv\'enic interactions and auroral emission from the brown dwarf (Equations~\ref{eq:power sub alfvenic} and \ref{eq:power aurora}).

\clearpage
\section{Radial Velocities, Priors, and Joint Sampling Posteriors}\label{sec:appendix_rvs}
Figure 10 shows the corner plot of the posterior distributions of the joint Gaia and RV sampling with the scaled Gaia covariance matrix ($\sigma_{\rm{scale}}$>1) from Section 4.3. Table 4 lists the dates, radial velocities, and associated errors for the RVs from TRES and HPF. Table 5 lists the priors used in the RV-only, Gaia+RV without inflation, and Gaia+RV with inflation MCMC runs for reproducibility purposes. 

\begin{deluxetable}{llcc}[h]
\tablecaption{Radial Velocities for LHS 1610A
\label{tab:appendix_rvs}}
\tabletypesize{\scriptsize}
\tablehead{\colhead{BJD (days)} &  \colhead{RV ($\text{km}\:\text{s}^{-1}$)}     & \colhead{RV error ($\text{km}\:\text{s}^{-1}$)}                               & \colhead{Instrument}}
\startdata      
2457785.7131             &  28.448             & 0.028     & TRES \\
2457786.7850             &  32.365             & 0.028     & TRES \\ 
2457787.6378             &  35.502             & 0.028     & TRES \\ 
2457794.6483             &  22.514             & 0.028     & TRES \\ 
2457795.7182             &  26.224             & 0.028     & TRES \\ 
2457800.7416             &  44.533             & 0.028     & TRES \\ 
2457806.6698             &  27.585             & 0.028     & TRES \\ 
2457807.6875             &  31.293             & 0.029     & TRES \\ 
2457808.6590             &  34.944             & 0.028     & TRES \\ 
2457821.6194             &  43.586             & 0.028     & TRES \\
2457822.6458             &  45.893             & 0.029     & TRES \\ 
2457823.6552             &  40.479             & 0.031     & TRES \\ 
2457824.6210             &  25.451             & 0.029     & TRES \\
2459092.921787           & -0.9804             & 0.0036    & HPF  \\
2459157.934541           &  1.1535             & 0.0028    & HPF  \\
2459185.659386           & -10.612             & 0.003     & HPF  \\
2459212.786561           &  -21.375            & 0.006     & HPF  \\
2459274.599746           &  0.6958             & 0.0061    & HPF  \\
2459621.664956           &  -4.5927            & 0.0092    & HPF  \\
\enddata
\tablenotetext{}{TRES RVs are adopted as provided in \cite{Winters2018}.}
\end{deluxetable}

\begin{deluxetable*}{lccr}
\centering
\tablecaption{Priors for the RV-only and Gaia+RV sampling. Priors labeled $\mathcal{U}$ are uniform within those bounds. Those with $\mathcal{N}$ are normal priors with the first value being the mean and the second value being the standard deviation of the Gaussian distribution. Priors with only a numerical value are fixed at that value.}
\label{tab:priors}.
\tabletypesize{\scriptsize}
\tablehead{\colhead{Parameter}  &  \colhead{RV-only Prior} & \colhead{Gaia+RV Prior - No Inflation} & \colhead{Gaia+RV Prior - Inflation}}
\startdata
M$_*$ ($M_{\odot}$)    & -                                & $\mathcal{N}(0.1671, 0.0041)$  & $\mathcal{N}(0.1671, 0.0041)$   \\
m$_2$ ($M_{\rm Jup}$)  & -                                & $\mathcal{U}(1.0,100.0)$       & $\mathcal{U}(1.0,100.0)$  \\
$\cos i$               & -                                & $\mathcal{U}(-1.0,0.0)$        & $\mathcal{U}(-1.0,0.0)$ \\ 
i ($^\circ$)           & -                                & -                              & - \\
K (m/s)                & $\mathcal{U}(100000.,200000.)$   & -                              & - \\
e                      & $\mathcal{U}(0.0,0.9)$           & $\mathcal{U}(0.0,0.9)$         & $\mathcal{U}(0.0,0.9)$ \\
$\omega$ (degs)        & $\mathcal{U}(0.0,360.0)$         & $\mathcal{U}(0.0,360.0)$       & $\mathcal{U}(0.0,360.0)$ \\
$\Omega$ (degs)        & -                                & $\mathcal{U}(-180.0,180.0)$    & $\mathcal{U}(-180.0,180.0)$  \\
t$_{peri}$ (days)$^{a}$& $\mathcal{U}(-6.0,6.0)^a$        & $\mathcal{U}(-6.0,6.0)^a$      & $\mathcal{U}(-6.0,6.0)^a$   \\
P (days)               & $\mathcal{U}(10.5638,10.6198)$$^{b}$& $\mathcal{U}(10.5638,10.6198)$$^{b}$    & $\mathcal{U}(10.5638,10.6198)$$^{b}$    \\
$\varpi$ (mas)         & -                                & $\mathcal{N}(103.879, 0.023)$  & $\mathcal{N}(103.879, 0.023)$ \\
$\gamma_{TRES}$ (m/s)  & $\mathcal{U}(-2000.0,2000.0)$    & $\mathcal{U}(-2000.0,2000.0)$  & $\mathcal{U}(-2000.0,2000.0)$  \\
$\gamma_{HPF}$ (m/s)   & $\mathcal{U}(-17000.0,17000.0)$  & $\mathcal{U}(-17000.0,17000.0)$& $\mathcal{U}(-17000.0,17000.0)$ \\
$\sigma_{\text{scale}}$  & -                                & 1.0                            &  $\mathcal{U}(1.0,25.0)$ \\
$\varepsilon$          & -                                & 0                              & 0          \\
\enddata
\tablenotetext{}{$^a$For the periastron time, we follow the Gaia convention where the periastron time is $t_{\mathrm{peri}} = 2457389.0 + t_p$, where $t_p$ is the value listed in the table above.}
\tablenotetext{}{$^b$This prior is a $\pm$10 sigma window of the \cite{Winters2018} period of $10.5918 \pm 0.0028$ days. This encompasses the period from the Gaia two-body solution.}
\end{deluxetable*}

\begin{figure*}
\centering
\includegraphics[width=\textwidth]{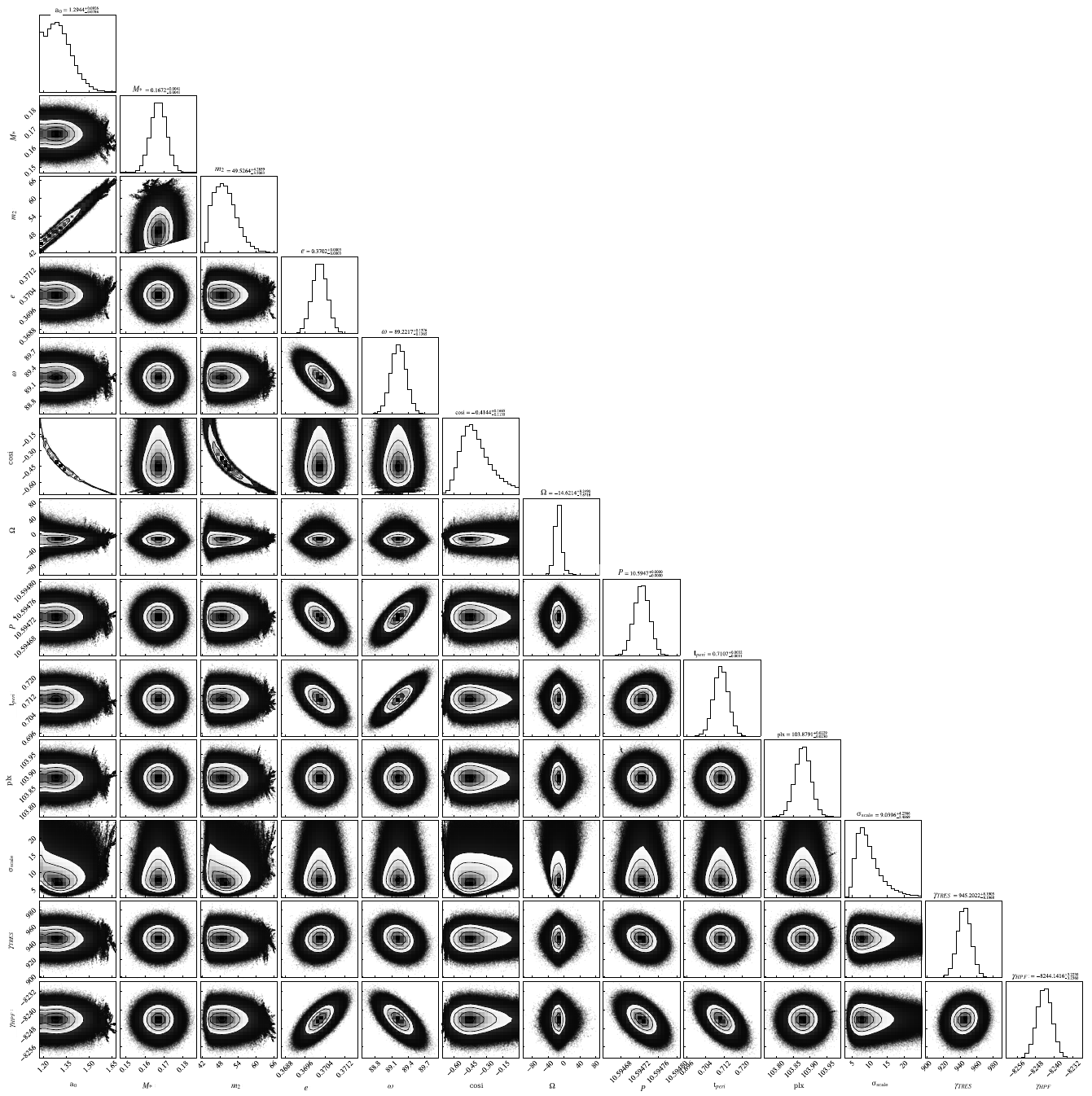}
\caption{Corner plot showing the posterior distributions of the joint Gaia and RV sampling with the scaled Gaia Covariance matrix ($\sigma_{\mathrm{scale}}>1$) from Section \ref{subsec:jointfit}. Median values are highlighted in Table \ref{tab:results}.}
\label{fig:corner_plot}
\end{figure*}

\clearpage
\bibliography{references.bib}{}
\bibliographystyle{aasjournal}

\end{document}